\documentclass[nofootinbib,twocolumn,aps,pre,superscriptaddress,citeautoscript,floatfix]{revtex4-1}
\usepackage{graphics}
\usepackage{graphicx}
\usepackage{color}
\usepackage{amsmath}
\usepackage{amssymb}
\usepackage{float}
\newcommand\be{\begin{equation}}
\newcommand\ee{\end{equation}}
\newcommand\bea{\begin{eqnarray}}
\newcommand\eea{\end{eqnarray}}
\begin{document}

\title{Orienting thin films of lamellar block copolymer:\\
the combined effect of mobile ions and electric field}
\author{Bin Zheng}
\affiliation{CAS Key Laboratory of Theoretical Physics, Institute of Theoretical Physics, Chinese
Academy of Sciences, Beijing 100190, China}
\author{Xingkun Man}
\email{manxk@buaa.edu.cn}
\affiliation{Center of Soft Matter Physics and its Applications, and School of Physics and Nuclear Energy Engineering, Beihang University, Beijing 100191, China}
\author{Zhong-can Ou-Yang}
\affiliation{CAS Key Laboratory of Theoretical Physics, Institute of Theoretical Physics, Chinese
Academy of Sciences,
Beijing 100190, China}
\author{M. Schick}
\affiliation{Department of Physics, University of Washington, Seattle, Washington 98195, USA}
\author{David Andelman}
\email{andelman@post.tau.ac.il}
\affiliation{Raymond and Beverly Sackler School of Physics and Astronomy, Tel Aviv University, Ramat Aviv 69978, Tel Aviv, Israel}


\begin{abstract}

We study thin films of A/B diblock copolymer in a lamellar phase confined between
two parallel plates (electrodes) that impose a constant voltage across the film.
The weak-segregation limit is explored via a Ginzburg-Landau-like free-energy expansion.
We focus on
the relative stability of parallel and perpendicular orientations of  the lamellar phase, and
how they are affected by variation of the following four experimental controllable parameters: addition of free ions,
the  difference in ionic solubilities
between the A and B blocks, their dielectric contrast, and the
preferential interaction energy of the plates with the blocks. It is found that, in general, the addition
of ions lowers the critical voltage needed to reorient the lamellae from being parallel to the
plates, to being perpendicular to them. The largest reduction in critical voltage is obtained
when the ions are preferentially solubilized in the block that is not preferred by the plates. This reduction is even further enhanced when the dielectric constant of this block has the higher value.
These predictions are all subject to experimental test.

\end{abstract}

\maketitle

\section{Introduction}

Block copolymers are polymeric systems composed of two or more
chemically distinct blocks,
covalently joined. They
self-assemble into structures with a characteristic scale on the order of a few to hundreds of nanometers~\cite{Fredrickson2006}.
The simplest block copolymer (BCP) system is that of a diblock copolymer,
in which each chain is composed of two blocks.
The well-studied phase behavior of
A/B BCP melts shows
various three-dimensional morphologies including lamellae,
hexagonally close-packed cylinders, body-centered cubic packing of spheres, and gyroid
networks~\cite{Matsen1994,Bates1994}. These bulk structures can be
controlled and adjusted by changing the three following parameters: the fraction,
$f$, of the A block in a chain, the Flory-Huggins $\chi$ parameter that
is related to the temperature, and the BCP chain length,
$N$~\cite{Semenov1985,Leibler1980,Fredrickson1987}.

Thin films of BCP have been intensively studied  because of
their significant potential for applications in several technologies, such
as those employed in the micro-electronics industry~\cite{Zschech2007,Stoykovich2005}.
For example, one can use nano-lithography techniques based on BCP
self-assembly to produce complex nano-materials. This is considered  a
promising strategy to confront the challenges of the next generation
computer-chip production~\cite{Park1997,Thurn2000}. However, an ever-present
difficulty that constrains its wider application, is the requirement to
produce perfectly aligned and defect-free self-assembled structures on
lateral length scales of dozens of microns or more.

To address those issues, a large body of
work~\cite{Amundson1991,Morkved1996,Olszowka2009,Majewski2012,Sivaniah2003,Kang2012,Man2012,Man2012a,Man2015,Man2016,Pujari2012,Zhang2014,Zhang2016,Amundson1993,Amundson1994}
has been devoted to
understanding and controlling the orientation of BCP thin films, and
the means by which the defects in their patterns can be eliminated. These
studies show that the structural requirements can be achieved by subjecting
the BCP films to a variety of external fields and treatments, such as
electric~\cite{Amundson1991,Amundson1993,Amundson1994,Morkved1996,Olszowka2009}
and magnetic~\cite{Majewski2012} fields, surface
patterning~\cite{Sivaniah2003,Kang2012,Man2012,Man2012a,Man2015,Man2016,Zhang2014,Zhang2016} and shear
forces~\cite{Pujari2012}.

In several experimental studies~\cite{Thurn-Albrecht2000,Xu2003,Xu2004,Wang2006}, external electric fields were used to
align and orient thin PS-b-PMMA  films with a cylindrical morphology.
In the absence of any electric field, the cylinders were oriented parallel to the film's substrate. Application of a voltage difference $V$,
across the film thickness creates a perpendicular electric field. Above some voltage threshold, $V_c$, full alignment of cylindrical domains oriented in a perpendicular direction to the substrate (and in the direction of the electric field) was achieved.

One way of reducing the value of $V_c$, is to reduce the substrate/film interactions by adsorbing onto the substrate a layer of random copolymer brushes, as was done in Ref.~\cite{Xu2003}.
In follow-up studies~\cite{Xu2004,Wang2006}, the influence of ionic impurities (such as LiCl) on PS-b-PMMA alignment has been investigated.
When the concentration of added LiCl was greater than a certain value (about 210 ppm), it was claimed that the hexagonal micro-domains can be fully oriented perpendicularly to the film surface,
regardless of the strength of interfacial interactions.
It was also suggested~\cite{Wang2006} that the Li$^+$ ions
complex selectively with the PMMA block, causing an increase of the PMMA
dielectric constant. Larger dielectric differences decrease the
$V_c$ value and, in general, enhances the
ability of the  electric field to orient thin BCP micro-domains.

Motivated by these experiments, several theoretical studies~\cite{Pereira1999,Ashok2001,Tsori2001,Tsori2002,Tsori2006,Lin2005,Tsori2006,Dehghan2015,Tsori2003,Putzel2010}
focused on the
orientation transition of BCP thin films subject to a constant voltage imposed by an external source.
The parameters used to model the film (beside $\chi$, $ N$ and $f$) are the
surface energies of the A and B blocks with the substrate, the
film thickness $L$, and the corresponding dielectric
constants, $\varepsilon_{\rm A}$ and $\varepsilon_{\rm B}$ of the two blocks. The theoretical methods
vary from a simple capacitor model of parallel and series stacking
of alternating A and B dielectric layers, applicable in the strong-segregation limit of large
$\chi N$~\cite{Pereira1999,Ashok2001}, to
Ginzburg-Landau (GL) expansions for the free
energy~\cite{Tsori2001,Tsori2002,Tsori2006} in the weak-segregation limit,
close to the Order-Disorder Temperature (ODT), $\chi_c N $, and to self-consistent field theories
(SCFT)~\cite{Lin2005,Tsori2006,Dehghan2015}.

The underlying physical mechanism for the orientation
transition has a simple origin. Initially, the lamellar phase is oriented parallel
to the two bounding surfaces,
because the surfaces have a preferential interaction with one of the two blocks.
Then, a voltage difference, $V$, is introduced across the film, and it
creates a perpendicular electric field of magnitude $E=V/L$, where $L$ is the film thickness.
Because the dielectric constants of the two blocks are different, the polarization of the system in the parallel (L$_{\parallel}$) and  perpendicular (L$_{\perp}$) orientations is different. In the latter L$_{\perp}$ orientation, the large polarization lowers the system free-energy (held at a fixed voltage by the external source).
Therefore, the electric field prefers a perpendicular orientation and competes with the surface interaction favoring the parallel phase. Eventually, as $V$ passes some critical value, $V_c$, the
system makes a transition from the  parallel orientation, L$_{\parallel}$, to the perpendicular one, L$_{\perp}$.
%

Given the above explanation for the reorientation transition in an electric
field, one would expect that adding free charges
that are preferentially solubilized in one of the blocks,
would enhance the polarizability of the L$_\perp$ orientation and,
thereby, lower the voltage, $V_c$, needed to bring about the transition~\cite{Tsori2003}.

A few theoretical works have studied this possibility.
Putzel {\it et al.}~\cite{Putzel2010} modeled the lamellar BCP thin film as a stack of alternating A/B
layers of two different dielectric constants ($\varepsilon_{\rm A}>\varepsilon_{\rm B}$).
The polymeric nature of the underlying blocks was not considered, and
the charges were assumed to be completely solubilized in one of the two blocks. Monovalent cations were fixed to
the A block and distributed uniformly. The anions were free to move and distributed themselves among the different
A lamellae, while they were not allowed to penetrate the B layers. It was found that the voltage $V_c$  was reduced by a small amount when electrical neutrality was
required for the entire  system. However, if each A lamella would satisfy individually  electro-neutrality, then
a large reduction in $V_c$  was found, due to the larger polarization of the perpendicular orientation as compared to the parallel one.

More recently, Dehghan {\it et al.}~\cite{Dehghan2015} revisited the effect of added ions in a BCP film. They considered the strong-segregation regime, $\chi N=20$, and utilized SCFT in which the polymer segment profiles and electric properties are statistical variables calculated
self-consistently.
Once again, the A blocks (with $\varepsilon_{\rm A}>\varepsilon_{\rm B}$) were taken to be uniformly charged by the anions and the cations could move freely throughout the system. The solubility of these mobile counter-ions in the B layers was taken to be much smaller than their solubility in the A layers. Two scenarios were investigated separately and resulted in opposite trends. When the B block (with the lower ionic solubility) is favored by the plates, the critical voltage $V_c$ decreased when the amount of free ions or  the difference in ion solubility between the two blocks increases.
However, when the A block (with the higher ionic solubility) is favored by the plates, the opposite occurs. Here, $V_c$ increases as a function of the amount of free ions and difference in ion solubility between the blocks.

In  the present study, and unlike previous works~\cite{Dehghan2015,Tsori2003,Putzel2010}, we consider a thin BCP film in the weak-segregation limit, $\chi N \gtrsim \chi_c N$. This means a more gradual A/B density profile, which leads to a different ionic distribution across the lamellar film.
Both cations and anions are mobile and can reach equilibrium by changing their local concentration.
The critical voltage, $V_c$, is calculated for all controllable system parameters, the surface free-energy between the  two bounding plates and the BCP film, the dielectric contrast and difference in ionic solubility of the A/B blocks, and the ion concentrations. Furthermore, we investigate separately the case in which the block with a favorable interaction with the plates is also the one with higher ionic solubility, and the opposite case in which it is not.

The analytic calculation is done in the framework of the Ginzburg-Landau free-energy expansion. The advantage of the relative simplicity of the calculation is that we are able to vary all parameters over substantial ranges. Although the concentration of free ions is in general small, the ionic effect can be rather large. From our analysis, a simple coherent picture emerges, permitting us to make several predictions about the dependence of $V_c$ on added ions and on several controllable system parameters. In particular, it was found that the addition of ions generally reduces the critical voltage needed to reorient the lamellae from the parallel to
perpendicular orientation.
%

The outline of the paper is as follows. In section~II, we introduce the model.
The equations for the spatial profiles of all relevant quantities are obtained by minimization of the  free energy in the two orientations, L$_\parallel$ and L$_\perp$.  In section~III, the relative
stability of the two phases is calculated numerically as function of the
applied voltage, $V$. The resulting critical voltage, $V_c$, is then obtained as a function of all relevant and controllable system parameters mentioned above, and done for two different cases:
(i)~the block with the {\it weaker} ionic solubility is favored by the plates (Figure~\ref{fig6}), and (ii)~the block with the {\it larger} ionic solubility is favored by the plates (Figure~\ref{fig7}). Finally, in section~IV, we isolate the various factors that contribute to the behavior of the critical voltage, $V_c$, and draw some conclusions (section~V) that are of value to future experiments.

\section{Model}

\begin{figure*}
{\includegraphics[width=0.85\textwidth,draft=false]{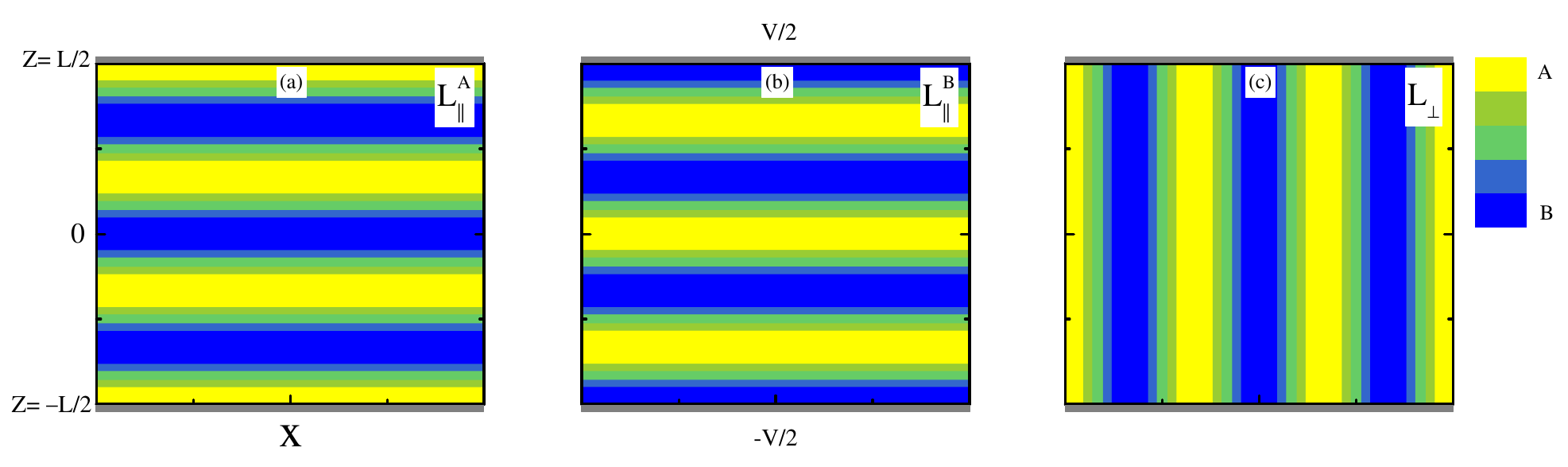}}
\caption{
\textsf{Schematic illustration of a lamellar block copolymer film in the $(X,Z)$ plane, confined between two flat and parallel surfaces located at $Z=\pm L/2$. The color code legend shows the relative A/B concentration in which yellow (gray) indicates A and blue (black) indicates B. In (a) and (b), the lamellar phases are oriented parallel to the surfaces and are denoted  L$_\parallel^{\rm A}$ and L$_\parallel^{\rm B}$, respectively. In (a), the A-monomers (yellow/gray) have a preferred surface interaction with the plates, while in (b), the B-monomers (blue/black) are preferred. In (c), the lamellae are in the perpendicular orientation, L$_\perp$. The distance between the two plates along the $Z$-direction is $L$. The two plates act as  electrodes imposing a voltage $V$ across the film, and no currents are allowed to flow in the system.
}}
\label{fig1}
\end{figure*}

We investigate the lamellar phase of A/B diblock copolymers in the weak-segregation limit,
which occurs close to the ODT  between the disordered and lamellar phases.
Each polymeric chain
contains $N=N_{\rm A}+N_{\rm B}$ monomers, $N_{\rm A}$ of the A block and $N_{\rm B}$
of the B block. The mole fraction of the two blocks is $ f=N_{\rm A}/N$ and $1-f=N_{\rm B}/N$, respectively.
As the two species are assumed to have the same monomeric volume, the A and B volume fractions are equal to
the corresponding A and B  molar fractions.

In a coarse-grained model, the local volume  fractions of the A and B monomers at position $\mathbf{R}=(X,Y,Z)$ are denoted, respectively, $\phi_{\rm A}(\mathbf{R})$ and $\phi_{\rm B}(\mathbf{R})$. Their spatial average is   $\langle \phi_{\rm A}\rangle=f$ and $\langle \phi_{\rm B}\rangle=1-f$. In the present study, only the symmetric BCP, $f=0.5$, is considered, although generalizations for $f\ne 0.5$ are straightforward and important,
for example, for the hexagonal phase.
For incompressible BCP melts, $\phi_{\rm A}(\mathbf{R})+\phi_{\rm B}(\mathbf{R})=1$ for all $\mathbf{R}$, and the thermodynamic quantities depend only on the relative volume (or molar) fraction, $\phi(\mathbf{R})\equiv \phi_{\rm A}(\mathbf{R})-\phi_{\rm B}(\mathbf{R})$, which serves as an order parameter.

In terms of the electrostatic properties, the A and B blocks are taken to be uncharged but  have different dielectric constants,
$\varepsilon_{\rm A}$ and $\varepsilon_{\rm B}$. Thus, the response of the layered BCP film to an external electric field is anisotropic.
It is reasonable to further assume a linear dependence of the local (coarse-grained) dielectric constant, $\varepsilon(\mathbf{R})$, on the local A/B concentration:
\be
\label{e1}
\varepsilon(\mathbf{R})=\varepsilon_{\rm A}\phi_{\rm A}(\mathbf{R})+\varepsilon_{\rm B}\phi_{\rm B}(\mathbf{R}) \, .
\ee
Expressed in terms of the relative BCP concentration, $\phi(\mathbf{R})$, this spatial dependence takes the form
\be
\label{e2}
\varepsilon(\mathbf{R})={\bar\varepsilon}+ \delta\varepsilon\,\phi(\mathbf{R})={\bar\varepsilon}[1+\lambda \phi(\mathbf{R})] \,
\ee
where
\bea
\label{e2.1}
{\bar\varepsilon}&=&(\varepsilon_{\rm A}+\varepsilon_{\rm B})/2 \, , \nonumber\\
\delta\varepsilon&=&(\varepsilon_{\rm A}-\varepsilon_{\rm B} )/2 \, , \nonumber\\
\lambda&=& \delta\varepsilon/{\bar\varepsilon} \, ,
\eea
where the parameter $\lambda$ is the relative dielectric contrast between the A and B blocks and will be used hereafter.

One of our main objectives  is to elucidate the effect of the relative  solubility of free ions in the A and B blocks on the voltage needed to reorient the film. This is done by the addition of monovalent cations and anions that are dissolved in the BCP melt. The solvation of the cations and anions depends on the local  relative concentration $\phi(\mathbf{R})$ of the A and B monomers, and their number density $n_\pm(\mathbf{R})$ is sufficiently small so that the ions do not affect the incompressibility condition of the pure BCP melt assumed above.

Figure~\ref{fig1} presents schematically the setup and the two considered lamellar orientations.
We will treat a BCP film of a lamellar morphology confined in the $(X,Z)$ plane. It is taken to be translationally invariant in the $Y$-direction, and confined between two flat and parallel
plates at $Z=\pm L/2$.

The two bounding plates have  multiple roles. They impose a rigid and planar boundary on the BCP melt, and they interact differently  with the A and B blocks. This preference is modeled by two surface interaction parameters $\tilde\sigma_{\rm t}$ for the top surface at $Z=L/2$, and $\tilde\sigma_{\rm b}$ for the bottom one at $Z=-L/2$. The plates, held at different fixed voltages, $\pm V/2$,  impose an electrostatic potential difference across the film. The capacitor-like anisotropic BCP film responds differently to the voltage in the parallel (L$_\parallel$) and perpendicular (L$_\perp$) orientations. This is manifested in a different ion profile in each of the two orientations (see, {\it e.g.}, Figures~\ref{fig2} -- \ref{fig4} of section~III). Those ionic profiles, in turn, interact with the anisotropic and heterogeneous dielectric profile of the film.

\subsection{Free energy}

The total free-energy of the lamellar BCP film with solvated free ions can be written as a sum of five terms discussed separately below,
\be
\label{e3}
F=F_{\rm pol}+F_{\rm p-i}+F_{\rm elec}+F_{\rm ion}+F_{\rm surf} \, .
\ee

The first term, $F_{\rm pol}$, is the Ginzburg-Landau (GL) expansion of the free energy of neutral BCP without any electrostatic effects.
It has been used extensively to describe lamellar and other BCP mesophases and their A/B density profiles~\cite{Fredrickson1987,Tsori2001,Tsori2002}.
This term can then be  written as
%
\be
\label{e3.1}
\begin{aligned}
\frac{F_{\rm pol}}{k_{\rm B}T\rho L^3}=\int {\rm d}^3r & \left[\frac{\tau}{2}\phi^2+
\frac{h}{2}(\nabla^2\phi+q_0^2\phi)^2
 +\frac{1}{24}\phi^4\right]\, ,
\end{aligned}
\ee
where hereafter the notation convention is that all lower-case lengths are rescaled by the film thickness $L$: $\mathbf{r}\equiv\mathbf{R}/L, x\equiv X/L, y\equiv Y/L$ and $z\equiv Z/L$.
The polymer density is $\rho=1/Nb^3$, where $b^3$ is the monomer volume.
The free-energy expression in eq~\ref{e3.1} is an expansion in powers of the
relative concentration, $\phi(\mathbf{r})=\phi_{\rm A}(\mathbf{r})-\phi_{\rm B}(\mathbf{r})$, and its spatial derivatives.
It depends on several phenomenological parameters. The first of these is the reduced temperature, $\tau=2 N(\chi_c-\chi)$, where $\chi$ is the Flory  parameter between the A and B monomers, with $N\chi_c\simeq{10.49}$ being the value at the critical point (ODT).
The second is $q_0$, which is the wavenumber, in units of $1/L$, of the lamellar phase. The third, $h$, is the dimensionless energy cost of spatial variations in the order parameter.
For further details on the choice of parameter values see Refs.~\cite{Fredrickson1987,Tsori2001,Tsori2002}

The second term, $F_{\rm p-i}$, is the non-electrostatic solvation energy of the cations and anions in the A/B melt. In general, this interaction (per $k_{\rm B}T$) can be written phenomenologically as
\be
\label{e4.1}
\left(\alpha_{+}^{\rm A}\,n_{+} + \alpha_{-}^{\rm A}\,n_{-}\right)\phi_{\rm A} + \left(\alpha_{+}^{\rm B}\,n_{+} + \alpha_{-}^{\rm B}\,n_{-}\right)\phi_{\rm B}  \, .
\ee
where $\alpha_{\pm}^{\rm A}$ and $\alpha_{\pm}^{\rm B}$ are the dimensionless ionic solvation parameters, and $n_{+}$ and $n_{-}$ are the number densities of positive and negative mobile ions.

In order to reduce the number of solvation parameters, a more restrictive case is considered, for which the cations and anions have equal solvation energy with the A (or B) blocks,
\bea
\label{e4.2}
\alpha_{\rm A}=\alpha_\pm^{\rm A} \, , \nonumber \\
\alpha_{\rm B}=\alpha_\pm^{\rm B} \, .
\eea
Then, the $F_{\rm p-i}$ term of the free energy reduces (up to a constant) to
\be
\label{e5}
\frac{F_{\rm p-i}}{k_{\rm B}TL^3}=\alpha\int{\rm d^3}r\,  (n_{-}+n_{+}) \phi \, ,
\ee
where

\be
\label{e5.1}
\alpha\equiv(\alpha_{\rm A}-\alpha_{\rm B})/{2}
\ee
is the only relevant ion-polymer solvation parameter in our study.
Note that throughout this work we take $\alpha$ to be {\it positive}, meaning that the ions solvate preferentially in the B block.

The third term, $F_{\rm elec}$, is the electrostatic energy, including the ion contribution. In SI units
\be
\label{e6}
\frac{F_{\rm elec}}{L^3}=
\int{\rm d^3}r \bigg[-\frac{\varepsilon_0\varepsilon({\bf r})}{2L^2}(\nabla \varPsi)^2
+e(n_{+} -n_{-})\varPsi\bigg],
\ee
where $\varPsi(\mathbf{r})$ is the electrostatic potential, $\varepsilon_0$ is the vacuum permittivity,
$e$ is the unit of electric charge, and $\varepsilon({\mathbf r})$ is the spatially-dependent dielectric constant introduced in eqs~\ref{e1}-\ref{e2.1}.

The fourth term, $F_{\rm ion}$, includes  the ion entropy of mixing
\be
\label{e7}
\frac{F_{\rm ion}}{k_{\rm B}TL^3}=\int{\rm d^3}r  \big[n_{-}\ln(a^3n_{-})\,+\,n_{+}\ln(a^3n_{+})
  -n_{-} - n_{+}\big]\, ,
\ee
with  $a^3$  being the volume
of the cations/anions, taken to be equal for both species.

The last term, $F_{\rm surf}$, is the surface interaction energy:
\be
\label{e8}
\frac{F_{\rm surf}}{L^2}=\int_{\rm S}{\rm d}^2r_s\,\tilde{\sigma}({\bf r}_s)\phi({\bf r}_s) \, ,
\ee
where the integration is over all bounding surfaces, and $\tilde{\sigma}$ is the surface  interaction coupled to the A/B surface concentration, $\phi(\mathbf{r}_s)$. For our two-plate system, the top and bottom plates have
a constant value of the surface interaction,
$\tilde{\sigma}_{\rm t}\equiv\tilde{\sigma}(z{=}0.5)$ and $\tilde\sigma_{\rm b}\equiv\tilde\sigma(z{=}-0.5)$.
Note that  ${\tilde \sigma}<0$ means that the plates prefer the A block, while for ${\tilde\sigma}>0$, they prefer the B block.

The BCP film is in contact with a
reservoir of BCP chains and ions. As the reservoir is electrically neutral, the number densities of anions and cations are both taken to be equal to $n_b$, the bulk salt concentration in the reservoir.

In order to proceed and calculate the various profiles, we need to minimize the total free energy with respect to $n_\pm$, $\varPsi$ and $\phi$. First, by minimization of the free energy with respect to $n_{\pm}$,  the sum and difference of $n_{\pm}$ can be expressed as:
\begin{eqnarray}\label{e9}
n_{+}+n_{-}&=&2n_{b}{\rm e}^{-\alpha\phi}\cosh( e\varPsi/k_{\rm B}T) \, , \nonumber\\
n_{+}-n_{-}&=&-2n_{b}{\rm e}^{-\alpha\phi}\sinh( e\varPsi/k_{\rm B}T).
\end{eqnarray}

It is convenient to introduce the following dimensionless quantities:
\be
\label{e12}
\begin{aligned}
&\psi({\bf r})=\frac{e\varPsi({\bf r})}{k_{\rm B}T}\, ,  &v=\frac{eV}{k_{\rm B}T}\, ,~~~~~    N_0&=\frac{n_b}{\rho} \\
&\sigma=\frac{{\tilde\sigma}}{k_{\rm B}T\rho L}\, , & \kappa_{\rm D}=L\left(\frac{2e^2n_b}{\varepsilon_0{\bar\varepsilon}k_{\rm B}T}\right)^{1/2}=\frac{L}{\lambda_{\rm D}}
\end{aligned}
\ee
where $N_0$ is the ratio between the ion and polymer bulk concentrations, and $\kappa_{\rm D}^{-1}=\lambda_{\rm D}/L$ is defined as the Debye screening length, $\lambda_{\rm D}$, in units of $L$, in a medium with dielectric constant $\bar\varepsilon$.

With the above definitions, the total free-energy of eq~\ref{e3}, with the use of eqs~\ref{e3.1}, \ref{e5} and \ref{e6}--\ref{e9}, takes the rescaled form
\be
\label{e12.1}
\begin{aligned}
\frac{F}{k_{\rm B}T\rho L^3}=&\int{\rm d}^3r \left\{\left[\frac{\tau}{2}\phi^2+\frac{h}{2}\left(\nabla^2\phi +q_0^2\phi\right)^2+
\frac{1}{24}\phi^4\right] \right.\\
&\left. -\frac{N_0}{\kappa_{\rm D}^2}(1+\lambda\phi)\left(\nabla \psi\right)^2-2N_0{\rm e}^{-\alpha\phi}\cosh\psi\right\} \\
 &+\int_{\rm S}{\rm d}^2r_s \,\sigma({\bf r}_s)\phi({\bf r}_s)\, .
\end{aligned}
\ee

The free-energy expressions for the
parallel (L$_\parallel$) and  perpendicular (L$_\perp$) lamellar orientations can now be written separately. The L$_\parallel$ phase can be further divided  into two sub-phases, L$_\parallel^{\rm A}$ and L$_\parallel^{\rm B}$, depending  on whether the A-rich lamellae (arbitrarily chosen to have the smaller ionic solubility), have a preferred interaction with the plates, or the B-rich lamellae are the preferred ones. See Figure~\ref{fig1} for an illustration of the three lamellar orientations.

\subsection{The  L$_\parallel$ phase}

For lamellae that are parallel to  two
infinite surfaces (electrodes) at $z=\pm 0.5$,  the system is translationally invariant
in the horizontal $x$ and $y$ directions. The lamellar free-energy density can be expressed from eq~\ref{e12.1} only as a function of $z$, and
for an $L\times L\times L$ film, the free energy $F_\parallel$ is written as
\be
\label{e19}
\begin{aligned}
\frac{F_\parallel}{k_{\rm B}T\rho L^3}=&\int{\rm d}z \left\{\left[\frac{\tau}{2}\phi^2+\frac{h}{2} \Bigl(\phi''(z) +q_0^2\phi\Bigr)^2+
\frac{1}{24}\phi^4\right] \right .\\
&\left . -\frac{N_0}{\kappa_{\rm D}^2}(1+\lambda\phi)\bigl[\psi'(z)\bigr]^2-2N_0{\rm e}^{-\alpha\phi}\cosh\psi\right\} \\
 &+\sigma_{\rm t}\phi_{\rm t}+\sigma_{\rm b}\phi_{\rm b}\, .
\end{aligned}
\ee
where $ {\rm d}\psi/{\rm d}{z}$ is denoted as $\psi'(z)$, etc.

Upon minimization of the above free-energy with respect to $\phi(z)$ and $\psi(z)$,
two coupled Euler-Lagrange equations are obtained for the A/B concentration, $\phi(z)$, and
the electrostatic potential, $\psi(z)$:
\be
\label{e13}
\begin{aligned}
h\phi''''+2hq_0^2\phi''+{\textstyle\frac{1}{6}}\phi^3+(\tau+ hq_0^4)\phi &\\
+\,2N_0\alpha{\rm e}^{-\alpha\phi}\cosh\psi-
\frac{\lambda N_0}{\kappa_{\rm D}^2}\psi^{'2}&=0\, ,
\end{aligned}
\ee
and
\be
\label{e14}
(1+\lambda\phi)\psi''+\lambda\phi'\psi'
-\kappa_{\rm D}^2{\rm e}^{-\alpha\phi}\sinh\psi=0\, ,
\ee
where eq~\ref{e14} is simply the Poisson-Boltzmann equation for a linear dielectric system with inhomogeneous $\varepsilon$.
In addition, the boundary conditions are:
\be
\label{e15}
\begin{aligned}
\phi''(\pm {\textstyle\frac{1}{2}})+q_0^2\phi(\pm {\textstyle\frac{1}{2}})&=0 \, , \\
\phi'''(\pm {\textstyle\frac{1}{2}})+q_0^2\phi'(\pm {\textstyle\frac{1}{2}})
\mp\frac{\sigma_{\rm t,b}}{h}&=0  \, ,\\
\psi(\pm {\textstyle\frac{1}{2}})&=\pm {\textstyle\frac{1}{2}}v \, ,
\end{aligned}
\ee
where $v\equiv eV/k_{\rm B}T$ is the dimensionless voltage imposed between the plates,
and $\sigma_{\rm t,b}$ are the top and bottom surface interactions, respectively.

\subsection{The L$_\perp$ phase}

The above derivation of the profile equations is repeated but this time for the  L$_\perp$ phase, in which the lamellae are  perpendicular to the two surfaces at $z=\pm 0.5$.
The A/B volume fraction, $\phi(x,z)$, changes most significantly along the $x$-axis, while the electrostatic potential
$\psi(x,z)$ changes most significantly along the $z$-axis.

The free energy $F$, eq~\ref{e12.1}, can be applied to the perpendicular orientation, and expanded to second order around its bulk value,
$F_\perp=F(\phi_0,\psi_0)+\delta F(\delta\phi,\delta\psi;\phi_0,\psi_0)$. As the expressions are somewhat cumbersome, they are presented in detail in the Appendix, and here we write the final expression for $F_\perp$ for an $L\times L\times L$ film
\be
\label{e20}
\begin{aligned}
&\frac{ F_\perp}{k_{\rm B}T\rho L^3}\\
&=\int{\rm d}x\int\, {\rm d}z \\
&\times \left\{\left[\frac{\tau}{2}\phi^2+\frac{h}{2}\left(\frac{\partial^2 \phi}{\partial x^2}+\frac{\partial^2 \phi}{\partial z^2}+q_0^2\phi\right)^2+\frac{1}{24}\phi^4\right] \right .\\
&\biggl.-\,\frac{N_0}{\kappa_{\rm D}^2}(1+\lambda\phi)\left[\left(\frac{\partial\psi}{\partial x}\right)^2+\left(\frac{\partial \psi}{\partial z}\right)^2\right] -\, 2N_0{\rm e}^{-\alpha\phi}\cosh\psi\bigg\} \\
 &+\int{\rm d}x \left[\sigma_{\rm t}\phi_{\rm t}(x) + \sigma_{\rm b}\phi_{\rm b}(x) \right]\, .
\end{aligned}
\ee

Further analytical progress can be obtained by assuming a single-mode variation of the profiles in the $x$-direction, as can be justified close to the ODT. This is done by assuming the following forms,
\begin{eqnarray}
\label{e16}
\delta\phi(x,z)&=&\phi(x,z)-\phi_0(x)=w(z)+g(z)\cos(q_0x)\, ,\nonumber\\
\delta\psi(x,z)&=&\psi(x,z)-\psi_0(z)=f(z)+k(z)\cos(q_0x) \, ,\nonumber\\
\end{eqnarray}
where $w,g, f$ and $k$ are amplitude functions that vary slowly in the $z$ direction.

The zeroth-order terms in the expansion are:
\begin{eqnarray}
\label{e17}
\phi_0(x)&=&\phi_q\cos(q_0x) \, ,\nonumber\\
\psi_0(z)&=&\psi_q\sinh(\kappa_{\rm D} z) \, ,
\end{eqnarray}
where $\phi_0(x)$ represents the density variation of a perpendicular lamellar phase with one mode, $q=q_0$. The function $\psi_0(z)$ is the Debye-H\"uckel solution of the linearized Poisson-Boltzmann equation with an average $\bar{\varepsilon}$ satisfying the L$_\perp$ boundary conditions, $\psi_0(\pm 0.5)=\pm v/2$.
The amplitude parameters $\phi_q$ and $\psi_q$ are,
\begin{eqnarray}
\label{e18}
\phi_q&=&\sqrt{-8\tau_{\rm eff}}\nonumber \, ,\\
\tau_{\rm eff}&=&\tau-2N_0{\alpha}^2 \, ,\\
\psi_q&=&\frac{v}{2\sinh(\kappa_{\rm D}/2)} \nonumber \, .
\end{eqnarray}

Next, the trial solutions of eqs~\ref{e16}-\ref{e17} are substituted into the free-energy expression, eq~\ref{e20}.
Minimizing the above free energy with respect to the amplitude functions $w(z)$, $g(z)$,
$f(z)$ and $k(z)$, results in four coupled equations with their corresponding boundary conditions,  as is shown in eqs~\ref{A2}-\ref{A6} of the Appendix.

\begin{figure*}
{\includegraphics[width=0.7\textwidth,draft=false]{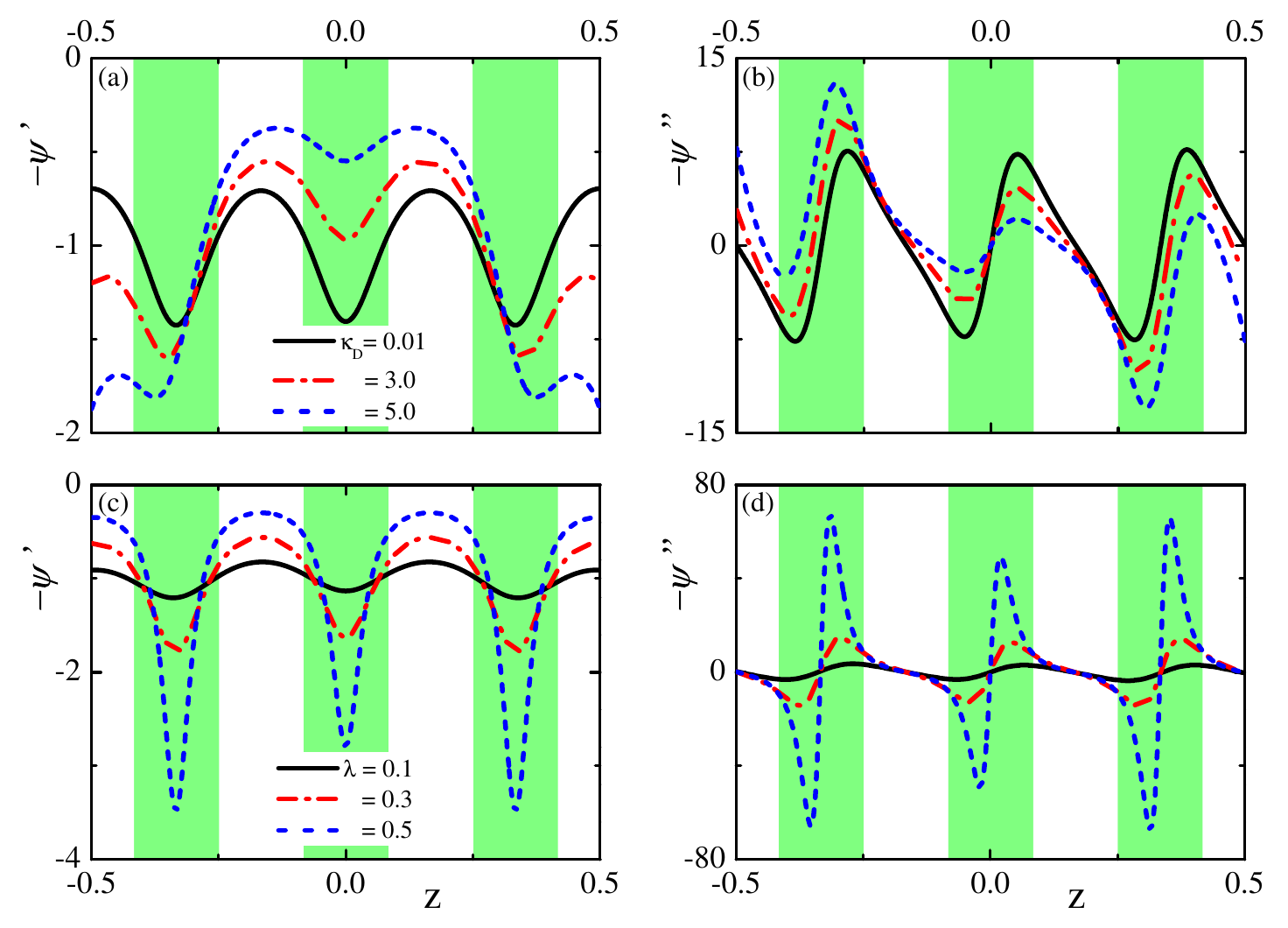}}
\caption{
\textsf{The electrostatic profiles for the parallel L$_\parallel^{\rm A}$ phase as function of the inter-plate distance, $-0.5\le z\le 0.5$. The two plates at $z=\pm 0.5$ are kept at a constant potential, $\pm v/2=\pm eV/(2k_{\rm B}T)$, and both prefer the A monomers ($\sigma<0$). The BCP has a periodicity $d_0=2\pi/q_0=1/3$ (in units of film thickness, $L$). For clarity, the A-rich regions ($\phi>0$) are marked as white and the B-rich ($\phi<0$) are colored green (gray). The dimensionless electric field, $E=-\psi'$, and local ion concentration, $\rho_c=-\psi''$, are plotted, respectively, in (a) and (b), in which $\kappa_{\rm D}$ (or equivalently $n_b$) varies as: $\kappa_{\rm D}=L/\lambda_{\rm D}=0.01$ ($n_b=5$\,nM, solid black line), 3.0 ($n_b=0.45$mM, dash-dotted red line)
and 5.0 ($n_b=1.25$mM, dashed blue line).
The other parameter values are: surface interaction $\sigma = -0.02$, ionic solubility $\alpha = 0.1$, dielectric contrast $\lambda = 0.2$ and $v = 1.0$. In (c)--(d), the effect of varying the dielectric contrast, $\lambda$, is shown for the values: $\lambda=0.1$ (solid black line), 0.3 (dash-dotted red line), 0.5 (dashed blue line). Other parameter values are $\sigma = -0.02$,  $\alpha = 0.1$, $\kappa_{\rm D} = 1.0$ and $v = 1.0$. The definition of all dimensionless parameters follows eq~\ref{e12}.
}}
\label{fig2}
\end{figure*}

\begin{figure*}
{\includegraphics[width=0.7\textwidth,draft=false]{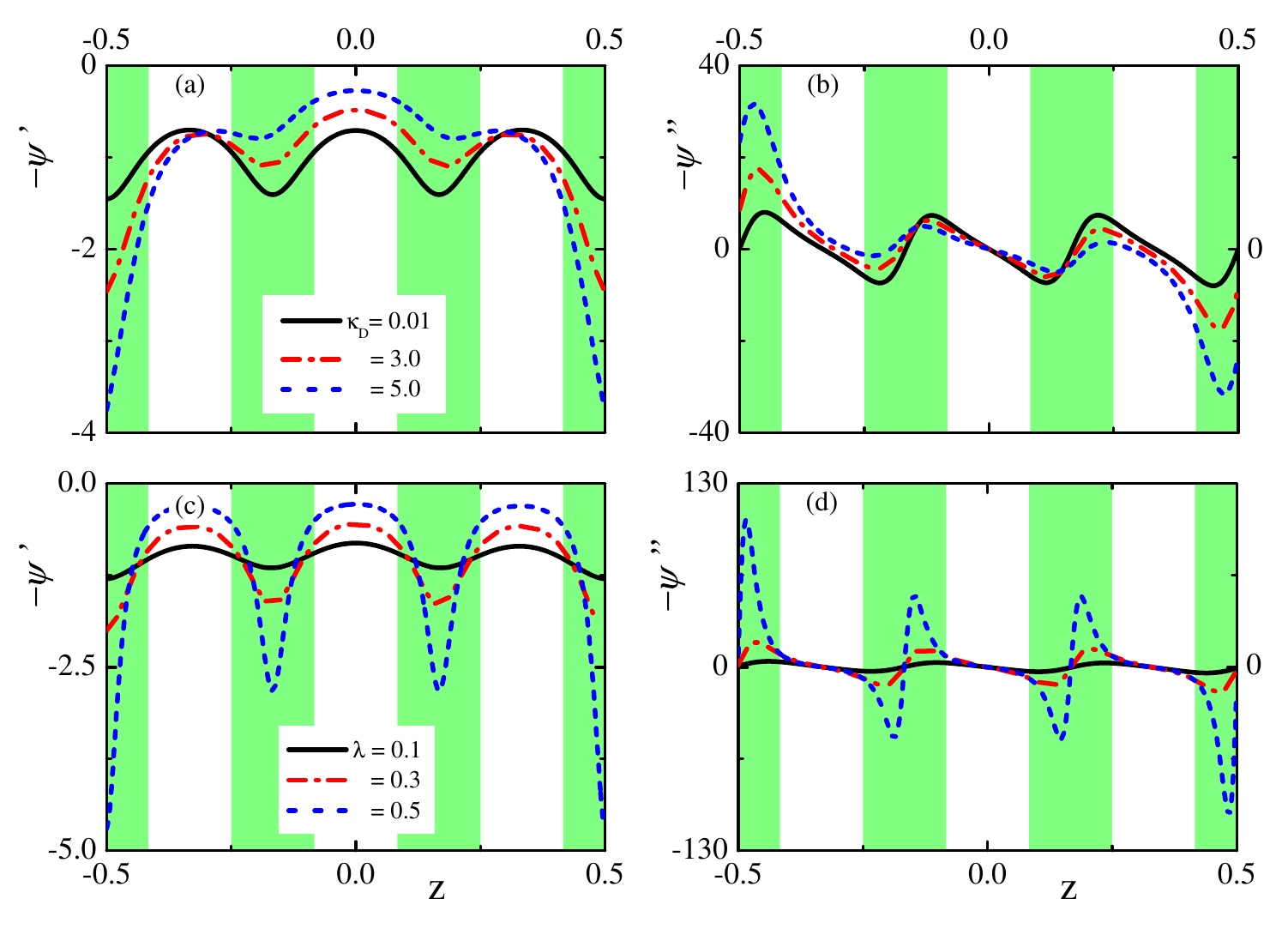}}
\caption{
\textsf{Same electrostatic profiles and parameters as in Figure~\ref{fig2}, but for the
parallel L$_\parallel^{\rm B}$ phase,
when both plates prefer the B monomers ($\sigma>0$).
The dimensionless electric field, $E=-\psi'$, and local ion concentration, $\rho_c=-\psi''$,
are plotted, respectively, in (a) and (b) for three values of $\kappa_{\rm D}=0.01$
($n_b=5$\,nM, solid black line), 3.0 ($n_b=0.45$mM, dash-dotted red line)
and 5.0 ($n_b=1.25$mM, dashed blue line). The other
parameters are: $\sigma = 0.02, \alpha = 0.1, \lambda = 0.2$ and $v = 1.0$.
In (c)--(d),  the same profiles are plotted as in (a)--(b), but
for three values of $\lambda=0.1$ (solid black line), 0.3 (dash-dotted red line), 0.5
(dashed blue line). Other parameter values are: $\sigma = 0.02, \alpha = 0.1,
\kappa_{\rm D} = 1.0$ and $v = 1.0$.
The definition of all
dimensionless parameters follows eq~\ref{e12}.
}}
\label{fig3}
\end{figure*}

\section{Results}

The calculations are done  for a film of cross-sectional area $L\times L=L^2$, in which the wavelength of the lamellae is rescaled, as are all lengths, by $L$. The BCP periodicity is $d_0=2\pi/q_0=1/3$ (meaning three A/B lamellae within the film thickness), and
the two plates at $z=\pm 0.5$ act as electrodes that are kept at a constant voltage difference, $v=eV/k_{\rm B}T$. The two plates are taken to have the same surface interaction with the A/B blocks, $\sigma=\sigma_{\rm t}=\sigma_{\rm b}$, with $\sigma>0$ corresponds to a surface preference for the B block.

The two parallel phases, L$_\parallel^{\rm A}$ and L$_\parallel^{\rm B}$,  shown schematically in Figure~\ref{fig1}a and b, should be considered separately.  The  L$^{\rm A}_{\parallel}$ phase is defined such that the surfaces prefer the A monomers ($\sigma<0$), while for the L$^{\rm B}_{\parallel}$ phase, the surfaces prefer the B monomers ($\sigma>0$). We note that in the absence of ions, and for symmetric BCP chains as considered here ($f=0.5$), the free energies of the two phases are equal under the change of B and A and $\sigma\to -\sigma$. This symmetry is broken once
ions are introduced because the ionic solubility in the two blocks is taken to be different, as is parameterized by $\alpha=(\alpha_{\rm A}-\alpha_{\rm B})/2 \neq 0$, eq~\ref{e5.1}.

In our study, we arbitrary choose only  $\alpha>0$, meaning that ions are preferentially solubilized in the B-rich regions.
Therefore, in the L$^{\rm A}_{\parallel}$ phase (Figure~\ref{fig1}a), the A layer is in contact with the plates, while
the ions are preferentially solubilized in B layers (as $\alpha>0$), which are not in direct contact with the plates.
This should be compared with the L$^{\rm B}_{\parallel}$ phase (Figure~\ref{fig1}b), in which the ions are preferably solubilized in the same B layer (as $\alpha>0$), which is also in contact with the plates. Note the other case of $\alpha<0$ can be simply obtained by the mapping: $\alpha\to-\alpha$, $\sigma \to -\sigma$ and $\lambda \to -\lambda$, and exchanging the meaning of the A and B  monomers for the symmetric  BCP ($f=0.5$), as will be discussed below.

\subsection{Electric field and ion density profiles}

The profiles of the electric field and the
ion density are plotted in Figure~\ref{fig2} for the L$_\parallel^{\rm A}$
phase, and in Figure~\ref{fig3} for the L$_\parallel^{\rm B}$ one.
Note that in the weak-segregation regime, there is no sharp interface between the A-rich  and  B-rich regions.  Nevertheless, for presentation purpose, the A-rich  regions ($\phi>0$) are shown in the two figures in white and the B-rich regions ($\phi<0$) in green (gray).

In Figure~\ref{fig2}a and b, we present, respectively, the dimensionless electric field in the $z$-direction,
$E=-\psi'(z)$, and the dimensionless total charge
density, $\rho_c=-\psi''(z)$. They are shown in Figure~\ref{fig2}a and b for three values of $\kappa_{\rm D}=L/\lambda_{\rm D}$, the dimensionless ratio between the film thickness $L$ and the Debye screening length $\lambda_{\rm D}$ (defined in eq~\ref{e12}):
$\kappa_{\rm D}=0.01$ (vanishingly small ion concentration, solid black line), 3.0 (dash-dotted red line), and 5.0 (large ion concentration, dashed blue line).
Taking a typical film thickness of $L \simeq 12.56$\,nm, an average dielectric
constant $\bar{\varepsilon}=4.5$ and a temperature $T=430$\,K, we obtain the density of the ions, $n_b$, as a function of $\kappa_{\rm D}$ as $n_b\simeq 0.05\kappa_{\rm D}^2$\,mM.
Figure~\ref{fig2}a shows the $z$-variation in the dimensionless electric field,
$E=-\psi'$, for these three values of $\kappa_{\rm D}$. With
added ions, the general trend is that the magnitude of the electric field near the plates
increases, while it decreases in the center of the film.
Figure~\ref{fig2}b presents the total charge density, $\rho_c=-\psi''$. Note that for extremely small ion concentration
($n_b=5$\,nM),
the charges are almost completely due to the bound polarization charges.

Figure~\ref{fig2} also demonstrates
one of the most significant differences between weak and strong
segregation. For  strong segregation~\cite{Putzel2010}),
the bound charge appears as sheets of dipoles located at the sharp
boundaries between the A and B layers.
However, for weak segregation considered here, the bound charges of different
signs are distinctly separated from one another, and their maximum and minimum values
do not occur at the boundary between A-rich and B-rich regions. We recall that
the A/B relative concentration has an almost sinusoidal variation $\phi \sim \cos (q_0x)$, while the white and green (gray)
regions in Figures~\ref{fig2} and \ref{fig3}  represent
the A-rich ($\phi>0$) and B-rich ($\phi<0$) regions simply to guide the eye.

As the electric field is stronger in the B regions (having the smaller dielectric constant, $\lambda>0$)
than in the A regions, both maxima
and minima in the local charge occur within the B regions (Figure~\ref{fig2}b and d).
This is supported by the
fact that, in the absence of free charge and for a linear dielectric, the
density of bound charge is given by $-\varepsilon_0\nabla\varepsilon\cdot{\bf E}/\varepsilon$.
As the three $\kappa_{\rm D}$ curves in Figure~\ref{fig2}b are not so different from one another, it follows that
the amount of added ions is relatively small, as compared to the bound polarization charge.

Figure~\ref{fig2}c and d shows, in a manner similar to Figure~\ref{fig2}a and b, the behavior
of  the $z$-component of the electric-field and ion density profiles, but now for
three different values of the dielectric contrast (eq~\ref{e2.1}) between the two blocks;
$\lambda=(\varepsilon_{\rm A}-\varepsilon_{\rm B})/(\varepsilon_{\rm A}+\varepsilon_{\rm B})=0.1,$
(solid black line) 0.3 (dash-dotted red line), and 0.5 (dashed blue line). As $\lambda$ increases,
a  pronounced perturbation to the electric field and the total free charge distribution can
be seen in Figure~\ref{fig2}c and d. While the A-rich layers (white) contain
almost no free ions, the B-rich layers (green/gray) accommodate both cations
and anions. A strong separation of positive and negative charges occurs
within each B layer, resulting in an effective dipolar layer within each B layer. This is
clearly seen in Figure~\ref{fig2}d, where the polarization charge density strongly depends
on the dielectric contrast $\lambda$. Because the A block is chosen here to be in contact with
the two plates, the polarization produced by this charge separation
is not as large as it would be were the B layers in contact with the plates, as is discussed next.

We turn now to the results for the L$_\parallel^{\rm B}$ phase. In Figure~\ref{fig3}, the spatial dependency of the electric field, $E=-\psi'$, and the ion concentration, $\rho_c=-\psi''$, are presented.
They are shown, respectively,
for three values of the free ion concentration (related to
variation in $\kappa_{\rm D}$), in Figure~\ref{fig3}a and b, and for three values of the dielectric contrast, $\lambda$, in Figure~\ref{fig3}c and d. Because the ions are preferentially solvated in the B (green/gray) regions that are now in contact with the plates, the  ions tend to accumulate at the electrodes, as seen in Figure~\ref{fig3}b and d. This effect is even more pronounced when either the density of ions or the dielectric contrast increase.
Comparison of Figure~\ref{fig3}b and \ref{fig3}d with Figure~\ref{fig2}b and d makes it clear that the separation of charge, and hence the polarization, is much larger in the L$_\parallel^{\rm B}$ phase than in the L$_\parallel^{\rm A}$ one. Note  the different scales of the electric field $E$, and ion concentration $\rho_c$, as  one compares  L$_\parallel^{\rm A}$ of Figure~\ref{fig2} with  L$_\parallel^{\rm B}$ of Figure~\ref{fig3}.

\begin{figure}
{\includegraphics[width=0.35\textwidth,draft=false]{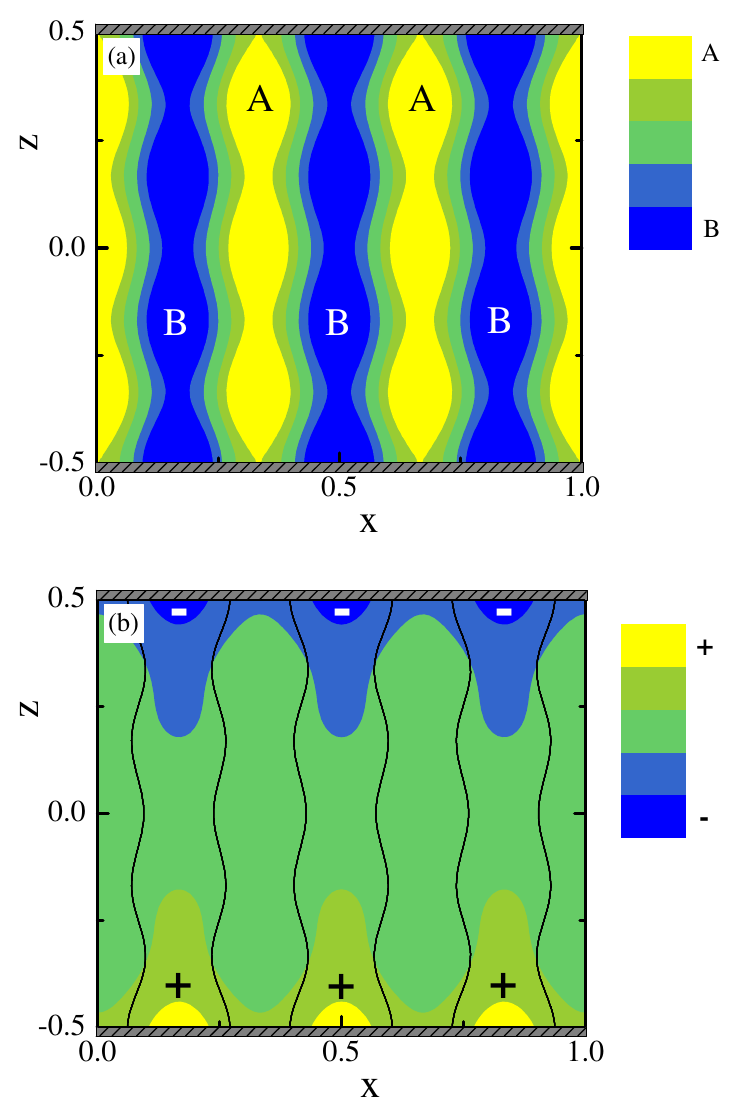}}
\caption{
\textsf{Contour plots in the $(x,z)$ plane for the L$_\perp$ phase, where the top and bottom
electrodes at $z=\pm 0.5$ are kept at a
constant voltage, $\pm v/2$.
(a) The A/B relative concentration $\phi(x,z)$, with the
A-rich regions as yellow (gray) and the B-rich ones in blue (black).
(b) Contour plot of the
local ion concentration, $n_{+}-n_{-}$, where the positive
charge density is colored in yellow (gray) and the negative one in blue (black). The thin black
lines in (b) are the A/B inter-material dividing surfaces (IMDS) defined
by $\phi = 0$. The parameter values in (a) and (b) are:
$\sigma=0.04, \kappa_{\rm D}=3.0, \alpha=0.4$ and $v=1.0$. From (a)  we see that with this value of $\sigma$,
the surface interaction is  strong enough for  the B monomers to almost completely cover the plates. In (b), the
cations (yellow/gray) accumulate in patches at the bottom
electrode, and a similar accumulation of anions (blue/black) is found at the top
electrode.
}}
\label{fig4}
\end{figure}

The perpendicular lamellar phase, L$_{\perp}$, is shown in Figure~\ref{fig4}a, where the contour plot of the monomer relative concentration, $\phi(x,z)$,  in the L$_{\perp}$ phase. The B block preferentially absorbs on the two electrodes, ($\sigma>0)$, while recalling that throughout our study we set (arbitrarily) $\alpha>0$, meaning that the ions prefers to be solvated in the B blocks. The interaction with the plates is sufficiently strong that the B monomers almost completely cover the plates.
As a consequence,  the A (yellow/gray) and B (blue/black) lamellae have a shape modulation rather than straight boundaries perpendicular to the electrodes.

Figure~\ref{fig4}b shows the contour plot of the local ion concentration. The black lines are the A/B inter-material dividing surfaces  defined as $\phi=0$. Clearly, the cations (yellow/gray) locally accumulate  at the negative bottom electrode, and similarly, the anions (blue/black) accumulate at the top positive one.
This leads to a charge separation on the order of the film thickness, $L$, as a result of the combined effect of surface preference for the B monomers ($\sigma>0$), and large solvation of cations/anions in the
B-rich regions (large $\alpha>0$).

\subsection{L$_\parallel$--to--L$_\perp$ transition in presence of ions}

\begin{figure}
{\includegraphics[width=0.45\textwidth,draft=false]{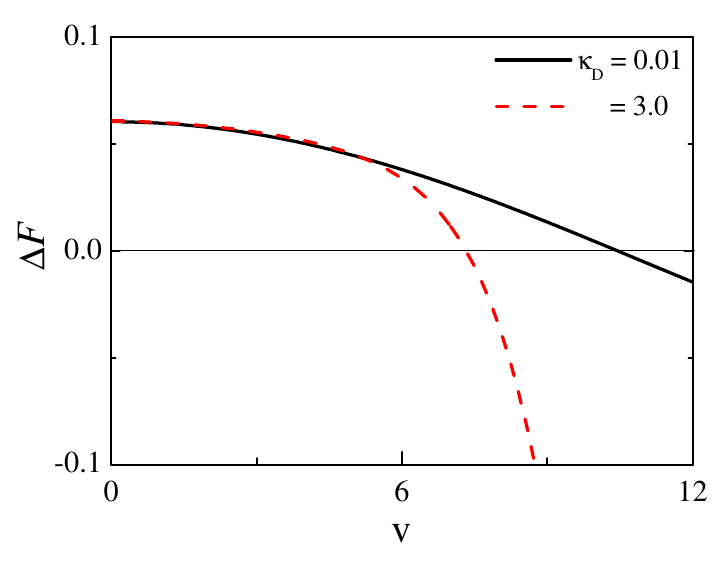}}
\caption{
\textsf{The free-energy difference, $\Delta F=F_\perp-F_\parallel^{\rm A}$,  between the  L$_\parallel^{\rm A}$ and  L$_\perp$ phases is plotted as function of $v$ for $\sigma=-0.02$ preferring the L$_\parallel^{\rm A}$ phase. Two ion concentrations are used and correspond to $\kappa_{\rm D}=0.01$ (solid black line) and  3.0 (dashed red line). The transition voltage $v_c$ is reduced as $\kappa_{\rm D}$ (ion concentration) increases; from $v_c\simeq 10.45$ for $\kappa_{\rm D}=0.01$ to $v_c\simeq 7.34$ for $\kappa_{\rm D}=3.0$. Other
used parameters are $\alpha=0.1$ and $\lambda=0.2$.
}}
\label{fig5}
\end{figure}

The orientation transition between the parallel to perpendicular orientations is investigated with special emphasis to the dependence of the critical voltage, $v_c$, on added ions.
We calculate the free-energy difference, $\Delta F=F_\perp-F_\parallel^{\rm A}$, between the perpendicular L$_\perp$ phase and the parallel L$_\parallel^{\rm A}$ one, in which the A blocks are preferred by the plates, as a function of the imposed external voltage $v\equiv eV/k_{\rm B}T$. One can see in Figure~\ref{fig5} that $\Delta F$ is a decreasing function of $v$.  It passes through zero at $v=v_c$, and  for $v>v_c$, the perpendicular orientation, L$_\perp$, is the globally stable phase.

Figure~\ref{fig5} compares $\Delta F$ for two different ion concentrations characterized by $\kappa_{\rm D}=0.01$ (almost no ions) and $3.0$.
The critical voltage, $v_c$, for the L$_\parallel^{\rm A}$--to--L$_\perp$ transition decreases with increasing density of free ions, $n_b$. The reduction in $v_c$ seen here is about 30$\%$. However,  by tuning the other system parameters, it is  possible to reduce $v_c$ by a factor of two purely by adding  a relatively small amount of free ions.
In a similar fashion, the free energy of the L$_\perp$ phase is compared  with that of the
parallel L$^{\rm B}_{\parallel}$ one, where  the B block is preferred by
the plates. We do not present these results here, but a similar reduction of $v_c$ is found for the same range of system parameters. A global view of the trends of $v_c$ with all four system parameters is presented next and in Figures~\ref{fig6} and \ref{fig7}.

\subsection{Trends of the critical voltage $v_c$}

\begin{figure*}
{\includegraphics[width=0.7\textwidth,draft=false]{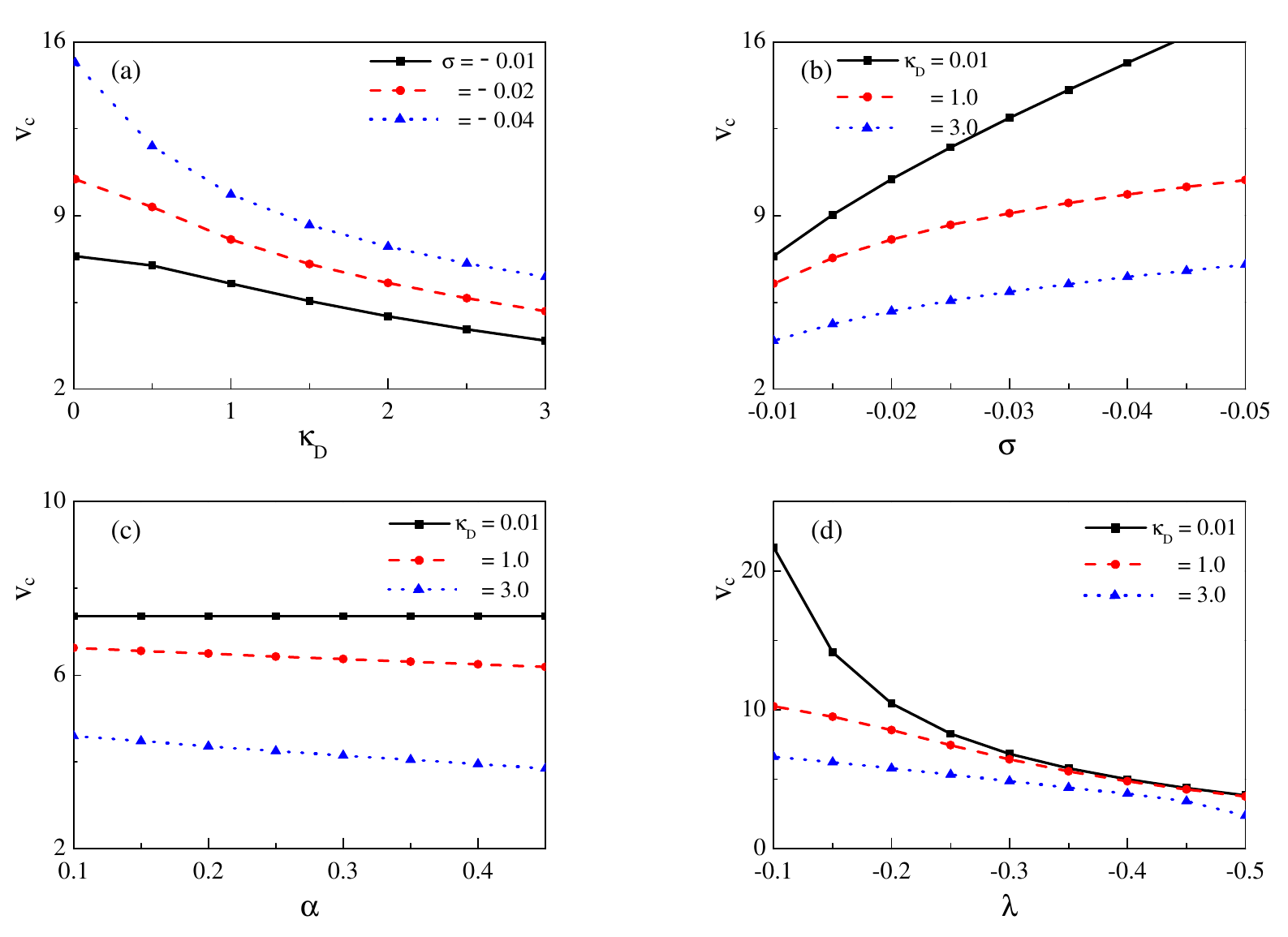}}
\caption{
\textsf{The phase diagram for the L$_\parallel^{\rm A}$--to--L$_\perp$ transition as function of four system parameters.
(a)~Critical voltage, $v_c$, as function of $\kappa_{\rm D}\sim \sqrt{n_b}$ for three $\sigma<0$ values, and for $\alpha=0.4$ and $\lambda=-0.2$. (b)~$v_c$ as function of surface-monomer interaction, $\sigma<0$, for three  $\kappa_{\rm D}$, and for $\alpha=0.4$ and $\lambda=-0.2$. (c)~$v_c$ as function of ion solvation parameter, $\alpha$, for three $\kappa_{\rm D}$, and for $\sigma=-0.01$ and $\lambda=-0.2$.
(d)~$v_c$ as function of the dielectric difference,
$\lambda\sim \varepsilon_{\rm A}-\varepsilon_{\rm B}$, for three different values of $\kappa_{\rm D}$, and for $\sigma=-0.02$ and $\alpha=0.1$.
}}
\label{fig6}
\end{figure*}

\begin{figure*}
{\includegraphics[width=0.7\textwidth,draft=false]{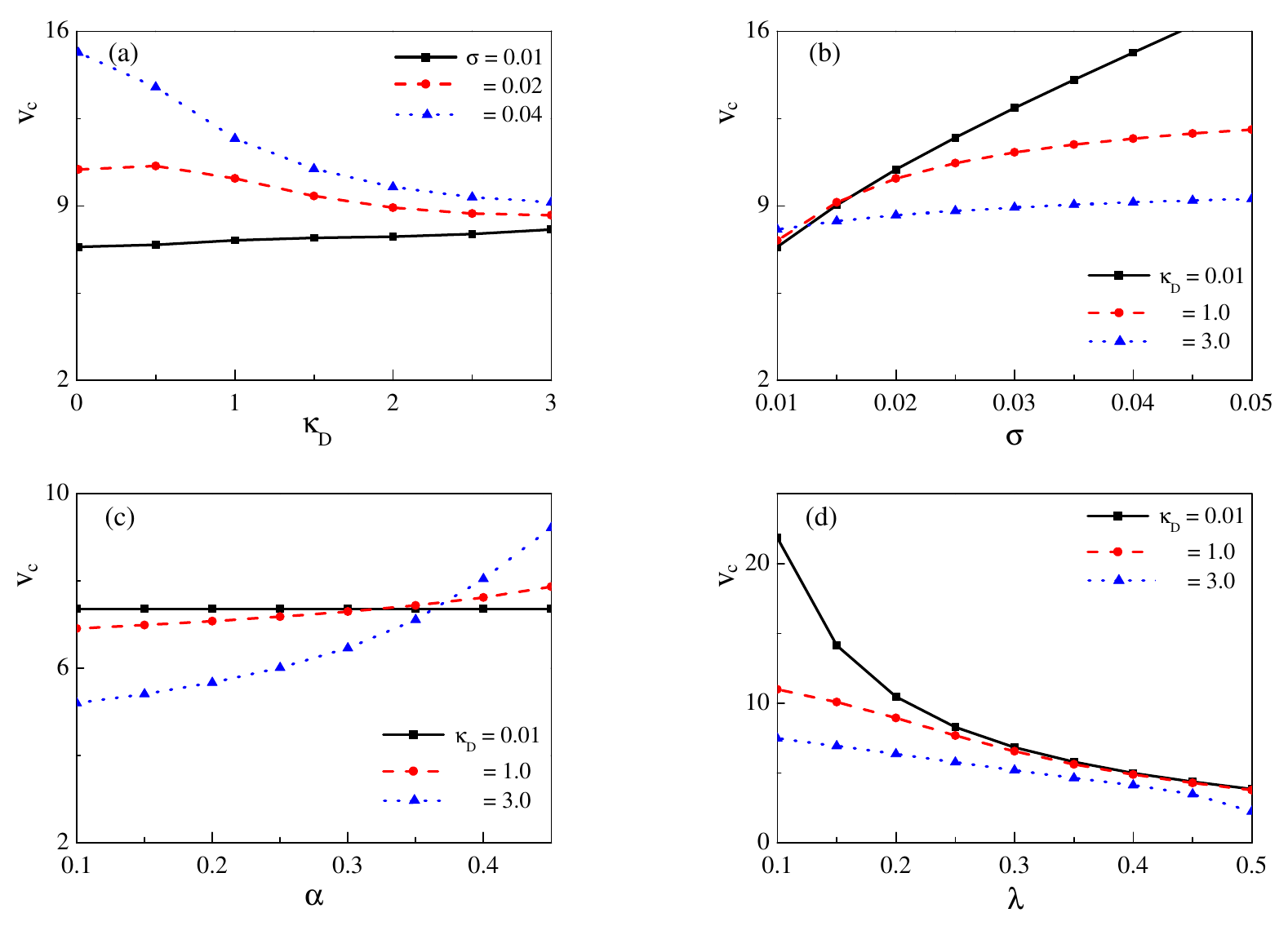}}
\caption{\textsf{
The phase diagram, similar to Figure~\ref{fig6} but for the  L$_\parallel^{\rm B}$--to--L$_\perp$ transition, plotted as function of four system parameters.
(a)~$v_c$ as function of $\kappa_{\rm D}\sim \sqrt{n_b}$ for three $\sigma>0$ values, and for  $\alpha=0.4$ and $\lambda=0.2$. (b)~$v_c$ as function of $\sigma>0$, for three $\kappa_{\rm D}$ values, and for  $\alpha=0.4$ and $\lambda=0.2$. (c)~$v_c$ as function of ion solvation parameter, $\alpha$, for three $\kappa_{\rm D}$ values, and for $\sigma=0.01$ and $\lambda=0.2$.
(d)~$v_c$ as function of the dielectric difference,
$\lambda\sim \varepsilon_{\rm A}-\varepsilon_{\rm B}$, for three $\kappa_{\rm D}$ values, and for $\sigma=0.02$ and $\alpha=0.1$.
}}
\label{fig7}
\end{figure*}

We show now results for the global dependence of the critical voltage, $v_c$, on several system parameters in Figure~\ref{fig6} for the L$_\parallel^{\rm A}$--to--L$_\perp$ transition (surfaces prefer A, $\sigma<0$), and in Figure~\ref{fig7} for the L$_\parallel^{\rm B}$--to--L$_\perp$ transition (surfaces prefer B, $\sigma>0$).
For the L$_\parallel^{\rm A}$ case, the dependence of $v_c$
on four parameters $\kappa_{\rm D}$, $\sigma$, $\alpha$ and $\lambda$, is shown separately in Figure~\ref{fig6}a-d. Each figure part contains three curves obtained for variation of one parameter while keeping the other two fixed.
The common trend is that $v_c$ decreases with the three parameters $\kappa_{\rm D}\sim \sqrt{n_b}$ (added ions), $\alpha$  (solvation contrast between A and B), and $\lambda$ (A/B dielectric contrast). Note that the reduction upon the addition of a small amount of ions can be quite large, on the order of 30\% to 50\%. As expected, the critical voltage increases with the surface parameter $|\sigma|$ as the latter prefers the L$_\parallel$ phase.

The second scenario of the trends of $v_c$ is analyzed for the L$_\parallel^{\rm B}$ phase, and is presented in Figure~\ref{fig7}. Unlike Figure~\ref{fig6}, here the surfaces as well as
the ions prefer the B block ($\sigma>0$), leading to accumulation of ions close to the plates. The increase of $v_c$ with
$\sigma$ (Figure~\ref{fig7}b) and its decrease with $\lambda$ (Figure~\ref{fig7}d) are similar to those shown in Figure~\ref{fig6}
above. However, the variation with $\kappa_{\rm D}$ and $\alpha$ is more complex and  left for further discussion in the next section.

\begin{figure}
{\includegraphics[width=0.45\textwidth,draft=false]{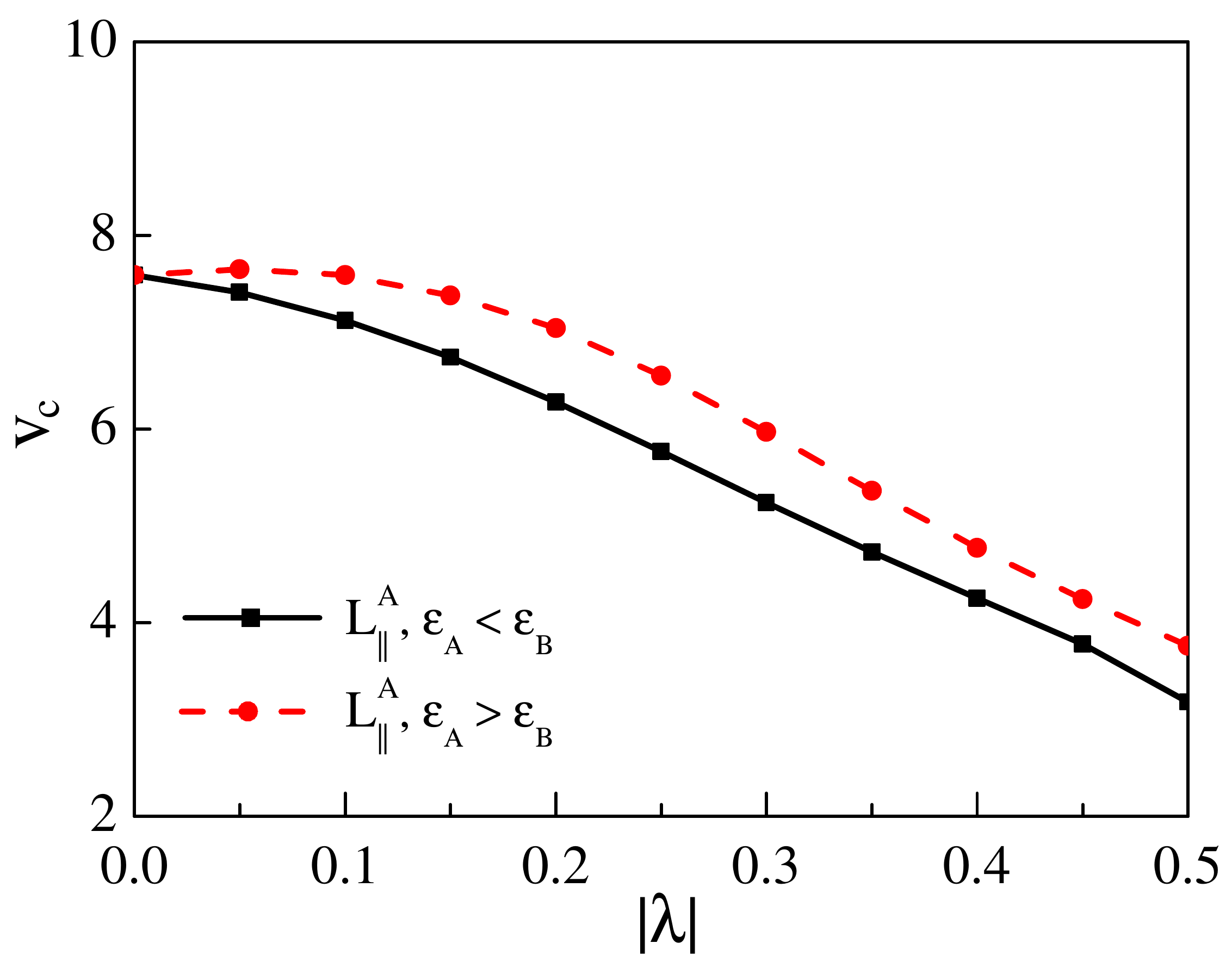}}
\caption{\textsf{Critical voltage, $v_c$, as a function of the magnitude of the dielectric
contrast, $|\lambda|$. The ions solubilize preferentially in the B-block,
$\alpha=0.4$.
The line of black squares shows $v_c$ for realignment of the L$_{\parallel}^{\rm
A}$ phase when the B-block has the larger dielectric constant ($\lambda<0$),
while the line of red circles represents $v_c$ for the same transition but when the
B-block has the smaller dielectric constant ($\lambda>0$). Other parameters are
$\kappa_{\rm D}=2.0$ and $\sigma=-0.02$.
}}
\label{fig8}
\end{figure}

\begin{table}
{\includegraphics[width=0.55\textwidth,draft=false]{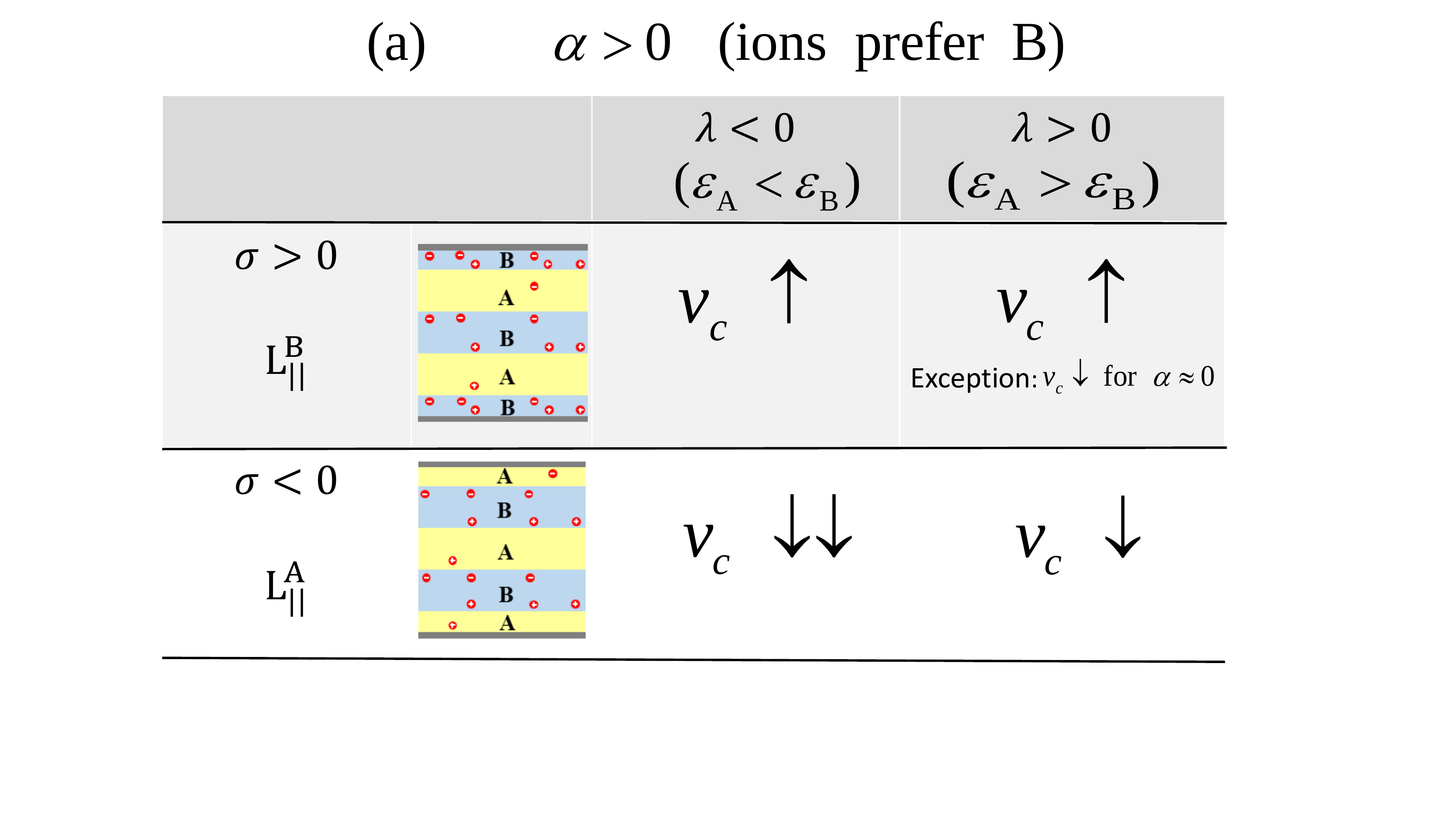}}
{\includegraphics[width=0.55\textwidth,draft=false]{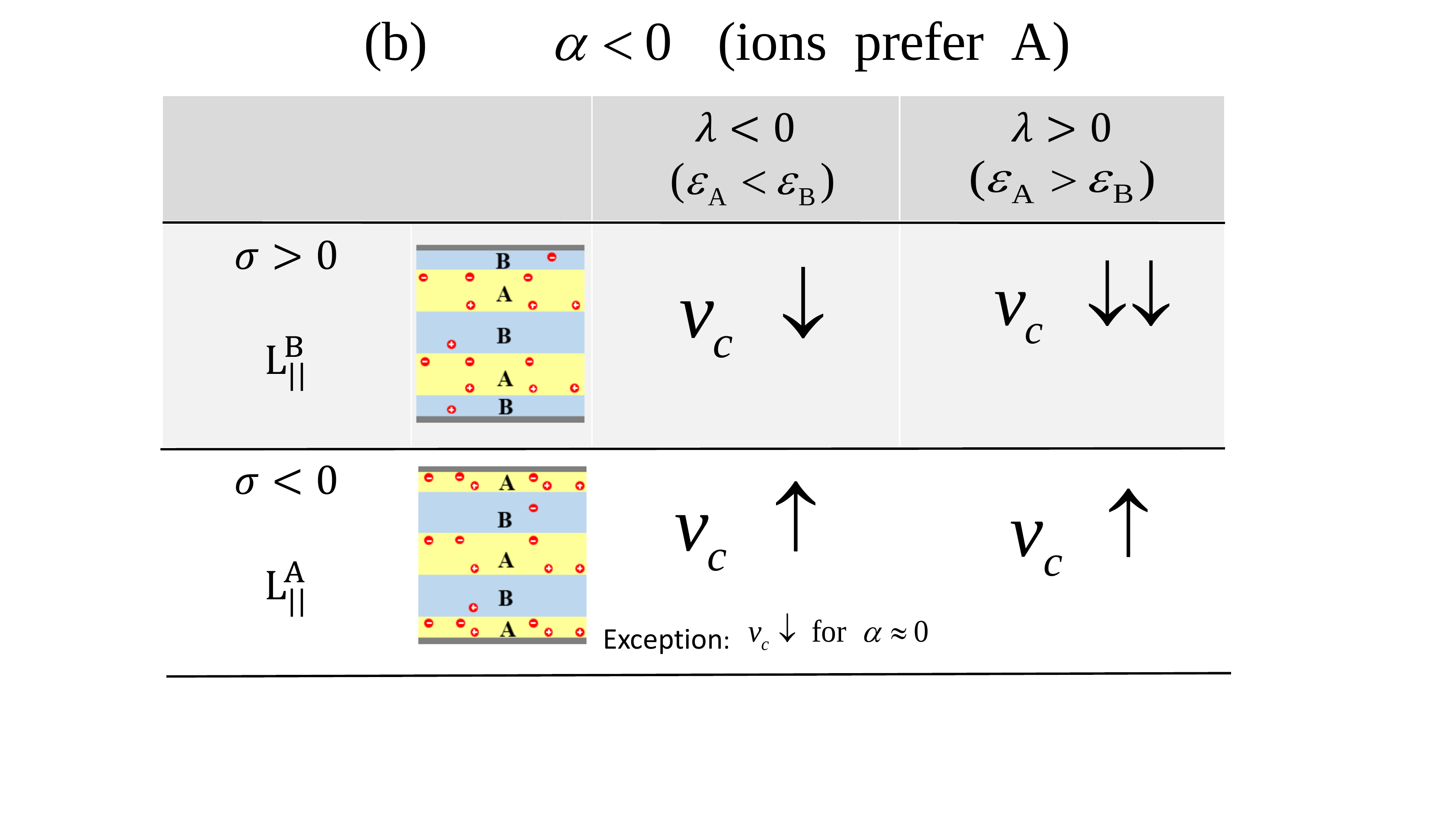}}
\caption{\textsf{The increasing or decreasing trends of $v_c$ for various combinations of the four key system parameters. (a) The various cases for $\alpha>0$, {\it i.e.,} the B monomers have a higher ionic solubility. The plates prefer either the A ($\sigma<0$) or B ($\sigma>0$) blocks. In addition, either the A block has the larger dielectric constant ($\varepsilon_{\rm A} > \varepsilon_{\rm B}$), or the B block has it ($\varepsilon_{\rm A} < \varepsilon_{\rm B}$).
 (b) The four above subcases are repeated but for $\alpha<0$, {\it i.e.,} the A monomers have a higher ionic solubility. Due to the fundamental symmetry, when we change $\alpha\to-\alpha$ as well as $\lambda\to-\lambda$ and $\sigma\to -\sigma$, the system with $f=0.5$ does not change its behavior.}
}
\label{table1}
\end{table}

As a final result, we present in Figure~\ref{fig8} the critical voltage $v_c$ as a function of $|\lambda|$ for the L$_\parallel^{\rm A}$ phase, and under fixed ion concentration ($\kappa_{\rm D}$) and $\sigma$. The line of black squares shows $v_c$  when the B block has the larger dielectric constant ($\varepsilon_{\rm A}<\varepsilon_{\rm B}$), while the line of red circles corresponds to $v_c$ when the B block has the smaller dielectric constant ($\varepsilon_{\rm A}>\varepsilon_{\rm B}$). Under the same conditions of ion concentration $n_b$ and surface interaction $\sigma$, it is clear from the figure that for the L$^{\rm A}_{\parallel}$ phase, smaller values of the applied voltage are sufficient to bring about
the reorientation, when the ions are preferably soluble in the block with the larger dielectric constant (in our example, $\varepsilon_A < \varepsilon_B $).

\section{Discussion}

The main aim of our study is to understand under what conditions the reduction of the critical voltage, $v_c$, can be optimized by adding free ions. We investigated in detail the system behavior as function of the following four adjustable and experimentally controlled parameters: the surface preference interaction with the A/B monomers, $\sigma$, the dielectric contrast $\lambda\sim \varepsilon_{\rm A}-\varepsilon_{\rm B}$, the ion solubility parameter $\alpha$, and, of course, the added-ion concentration $n_b\sim \kappa^{2}_{\rm D}$. The sign of $\alpha$ is arbitrarily chosen to be positive; namely, the ions prefer the B layers. The sign of $\sigma$ determines whether we deal with the parallel orientation for which the A layer is in contact with the plates (L$^{\rm A}_{\parallel}$ for $\sigma < 0$), or the B one (L$^{\rm B}_{\parallel}$ for $\sigma > 0$). Finally, the sign of $\lambda$ determines which of the two A/B layers has the higher dielectric constant.

In principle, one should study the dependence of $v_c$ on $n_b$ for all eight combinations of the three parameter signs. But there is a fundamental symmetry for a symmetric BCP ($f = 0.5$), where the volume fraction of the A and B monomers is equal. A simultaneous change of the sign of all three parameters $\sigma \rightarrow -\sigma$, $\lambda \rightarrow -\lambda$ and $\alpha \rightarrow -\alpha$, merely interchanges the labels A and B, and the system behaves exactly the same. Hence, we limit the study to $\alpha > 0$, and consider only the remaining four combinations of the sign of $\sigma$ and $\lambda$. This is presented in Table I.a, where we show for $\alpha>0$ under what conditions the value of $v_c$ will increase or decrease. Then, by using the above-mentioned symmetry, the other four different choices of signs for the $\alpha<0$ case are presented separately in Table I.b.

Figures~\ref{fig6} and \ref{fig7} show that for a giving set of experimental parameters ($\alpha$, $\sigma$, $\lambda$, $n_b$), $v_c$ for the L$^{\rm A}_{\parallel}$-to-L$_\perp$ transition is smaller than  for the L$^{\rm B}_{\parallel}$-to-L$_\perp$ transition in most cases. This can be simply understood as follows: As free ions are chosen to prefer the B-block ($\alpha>0$), the charge separation in the L$^{\rm B}_{\parallel}$ phase spans the entire film thickness, $L$, while it is one half periodicity less in L$^{\rm A}_{\parallel}$. This indicates that L$^{\rm B}_{\parallel}$ becomes thermodynamically more stable than L$^{\rm A}_{\parallel}$ in presence of additional free ions, as the dipolar contribution to the free energy is larger. Therefore, the external voltage needed to cause a transition from parallel to perpendicular orientation for L$^{\rm B}_{\parallel}$ is larger than for L$^{\rm A}_{\parallel}$. This effect can be seen by comparing Figs.~\ref{fig6} and \ref{fig7}, where $v_c$ is indeed larger for L$^{\rm B}_{\parallel}$ than L$^{\rm A}_{\parallel}$, while keeping all other system parameter at the same values. In Table I.a, this is shown by comparing the top row ($\sigma>0$) with the bottom one ($\sigma<0$), and indeed $v_c$ decreases more for the bottom row.

Since we have found that the critical field decreases more for the L$^{\rm A}_{\parallel}$ than for the L$^{\rm B}_{\parallel}$ phase (always keeping $\alpha>0$), it is enough to concentrate on the two remaining subcases for the L$_\parallel^{\rm A}$ phase appearing in the second row in Table I.a. They  differ  by the sign of $\lambda$, corresponding to whether $\varepsilon_{\rm A}$ is larger or smaller than $\varepsilon_{\rm B}$. Figure~\ref{fig8} compares those two cases and gives a clear answer that $v_c$ is smaller when $\varepsilon_B > \varepsilon_A$ ($\lambda<0$).
This result indicates that in order to decrease $v_c$, it would be more advantageous if two conditions are satisfied: (i) the block in which the ions are more soluble (here the B-block, $\alpha>0$), is also the block with the larger dielectric constant ($\lambda < 0$). (ii) This block is not in contact with the surface.

We now turn to the effects of adding free ions on $v_c$ for all four combinations of $\sigma$ and $\lambda$ for $\alpha>0$ (Table I.a). Our qualitative explanation is simple, although it does not retain the more complex coupling between the various parameters. In addition, it is in agreement with previous theoretical works~\cite{Dehghan2015,Tsori2003,Putzel2010}. As the charge separation for the L$^{\rm A}_{\parallel}$  is about the film thickness $L$ minus half a periodicity, while this separation spans the entire film thickness, $L$, for the $L_\perp$, it leads to a difference in polarization between the two phases; roughly, an effective charge multiplied by $d_0/2$. The addition of free ions decreases $\Delta F=F_\perp-F_\parallel$ because when the total charges increase, it will make the L$_\perp$ phase more stable. Therefore, $v_c$ decreases by adding additional free ions in L$^{\rm A}_{\parallel}$  for both $\lambda>0$ and $\lambda<0$ cases, as is seen in the second row of Table I.a.

On the other hand, the charge separation in the L$^{\rm B}_{\parallel}$ roughly spans the thickness $L$, just as it does for L$_\perp$. However, as the electrodes are completely covered by B blocks for the parallel phase, while they are less wetted by the B blocks for the perpendicular phase, the additional free ions makes the parallel phase more stable than the perpendicular one. This is why $v_c$ increases by increasing the free ion concentration, as is presented in the first row of Table I.a and in Fig~\ref{fig7}a  (black line, $\sigma=0.01$).

We remark that for the L$^{\rm B}_{\parallel}$ phase the situation is somewhat more complex than for the L$^{\rm A}_{\parallel}$ one. While for most parameter values, $v_c$ increases with $n_b$, as discussed above, there is an exception in this dependence as can be seen
in the top right entry of the Table I.a  (for $\lambda>0$).
The more usual case mentioned previously is valid for large $|\alpha|$ and small $\sigma$, and leads to an increase in $v_c$ with $n_b$. However, other combinations of $\alpha$ and $\sigma$  lead to a {\it decrease} of $v_c$ with $n_b$. As there are always some free ions in the A block, increasing $n_b$ enhances the stability of L$_\perp$ but has little effect on the stability of L$^{\rm B}_{\parallel}$. As a result, $v_c$ decreases for increasing values of $n_b$. This more subtle effect has not been mentioned in previous works.

\section{Conclusions}

To summarize, from our study that employs an analytic free-energy expansion in the weak-segregation limit, we can draw two important  conclusions: (i)~the addition of even a small amount of free ions generally reduces the critical voltage, $v_c$, needed to reorient the system of lamellae from parallel to perpendicular orientation with respect to the plates;
(ii) The largest reduction in $v_c$ is obtained when the ions are most soluble in a block which has the largest dielectric constant, and is also the block that is not preferred by the plates.
%
This is shown in Table I.a for the L$^{\rm A}_{\parallel}$ phase when we satisfy the conditions: $ \sigma<0, \alpha>0$  and  $\lambda<0$. We show, separately in Table~I.b that such conditions are satisfied for the L$^{\rm B}_{\parallel}$ phase for  $\sigma>0, \alpha<0$ and $ \lambda>0$.

These results are in line with several previous works but  offer a broader viewpoint as we consider in detail the combination of all important system parameters, and their effect on $v_c$.
These conclusions can certainly be tested in future experiments. In addition, it will be of interest to complement our analytical results by numerical works that are not restricted to the weak-segregation limit, and potentially to other anisotropic phases such as non-symmetric lamellar or  hexagonal BCP phases.

\bigskip
{\it Acknowledgment.}~~
We thank A. Cohen, W.-H. Mu and F.-F. Ye for useful discussions,
and, in particular, Y. Tsori for his instrumental help and suggestions in the initial stage of this work.
This work was supported in part by Grant No.~ 21434001 and 21404003 of the National Natural Science
Foundation of China (NSFC), the joint NSFC-ISF
Research Program, jointly funded by the NSFC under Grant
No.~51561145002 and the Israel Science Foundation (ISF)
under Grant No.~885/15, and the United States-Israel Binational Science Foundation (BSF)
under Grant No.~2012/060. D.A. acknowledges the hospitality of
the ITP (CAS) and Beihang University, Beijing, China, and
a CAS President's International
Fellowship Initiative (PIFI).

\appendix
\section{The free energy and profile equations of L$_\perp$}

As explained in section~II.C, the L$_\perp$ free energy of eq~\ref{e20} is expanded to second order around its bulk value, $F_\perp=F_\perp(\phi_0,\psi_0)+\delta F_\perp(\delta\phi,\delta\psi;\phi_0,\psi_0)$, yielding,

\be
\label{A1}
\begin{aligned}
&\frac{1}{k_{\rm B}TL^3\rho}\delta F_\perp(\delta\phi,\delta\psi;\phi_0,\psi_0)\\
&=\int{\rm d}x\int{\rm d}z \left\{\left[(\tau+hq_0^4)\phi_0+
\frac{1}{6}\phi_0^3+hq_0^2\nabla^2\phi_0 \right. \right. \\
&\left. +\, 2N_0\alpha{\rm e}^{-\alpha\phi_0}\cosh(\psi_0) -
\frac{N_0\lambda}{\kappa_{\rm D}^2}(\nabla\psi_0)^2\right]\delta\phi \\
&-\,2N_{0}{\rm e}^{-\alpha\phi_0}\sinh(\psi_0)
\delta\psi-\frac{2N_0(1+\lambda\phi_0)}{\kappa_{\rm D}^2}\nabla\psi_0\nabla\delta\psi \\
&+h(\nabla^2\phi_0+q_0^2\phi_0)\nabla^2\delta\phi \\
&+\frac{1}{2}\left[\tau+hq_0^4+\frac{1}{2}\phi_0^2
-2N_0{\alpha}^2{\rm e}^{-\alpha\phi_0}\cosh(\psi_0)
\right](\delta\phi)^2 \\
&+\frac{1}{2}h(\nabla^2\delta\phi)^2-
N_0{\rm e}^{-\alpha\phi_0}\cosh(\psi_0)
(\delta\psi)^2 \\
&-\frac{N_0(1+\lambda\phi_0)}{\kappa_{\rm D}^2}(\nabla\delta\psi)^2
+hq_0^2\delta\phi\nabla^2\delta\phi \\
&\left.
+ \,2N_0\alpha {\rm e}^{-\alpha\phi_0}\sinh(\psi_0)
\delta\phi\delta\psi-\frac{2N_0\lambda}{\kappa_{\rm D}^2}\nabla\psi_0\delta\phi\nabla\delta\psi\right\} \\
&+\int_{\rm S}{\rm d}x\,\sigma\delta\phi \, , \\
\end{aligned}
\ee
where $\psi=\psi_0+\delta\psi$, $\phi=\phi_0+\delta\phi$,
and the gradient operator is taken in the plane, $\nabla=({\partial}/{\partial x} ,{\partial}/{\partial z})$

The 2D profile equations for $\phi(x,z)$ and $\psi(x,z)$ of the L$_\perp$ phase are obtained by solving the four coupled differential equations for the amplitude functions, $w(z)$, $f(z)$, $g(z)$ and $k(z)$:
\be
\label{A2}
\begin{aligned}
&w''''+2q_0^2w''+\Bigl(q_0^4-\frac{{\tau}_{\rm eff}}{h}\Bigr)w \\
&=-\frac{2N_0\alpha}{ h}+\lambda\psi_q\frac{N_0}{\kappa_{\rm D} h}
\cosh(\kappa_{\rm D} z)\left[\psi_q\kappa_{\rm D}\cosh(\kappa_{\rm D} z) + 2f'\right] \\
&-\frac{2N_0\alpha}{h}\psi_q\sinh(\kappa_{\rm D} z)f -\frac{{\alpha}^3N_0}{h}\phi_q g \\
&+\frac{{\alpha}^2 N_0}{h}\phi_q\psi_q \sinh(\kappa_{\rm D} z)k
\end{aligned}
\ee

\be
\label{A3}
\begin{aligned}
&g''''-\frac{2{\tau}_{\rm eff}}{h}g\\
&=\frac{2\lambda\psi_q N_0}
{h\kappa_{\rm D}}\cosh(\kappa_{\rm D} z)k'
+\frac{2 N_0{\alpha}^2}{h}\phi_q\psi_q\sinh(\kappa_{\rm D} z)f \\
&-\frac{2 N_0\alpha}{h}\psi_q \sinh(\kappa_{\rm D} z)k
-\frac{2N_0{\alpha}^3}{h}\phi_q w
\end{aligned}
\ee

\be
\label{A4}
\begin{aligned}
&k''-(\kappa_{\rm D}^2+q_0^2)k \\
&=-\lambda\psi_q\kappa_{\rm D}\cosh(\kappa_{\rm D} z)g'-(\lambda+\alpha)\psi_q\kappa_{\rm D}^2\sinh(\kappa_{\rm D} z)(\phi_q+g) \\
&-\lambda\phi_q f''-\alpha\phi_q\kappa_{\rm D}^2f+{\alpha}^2\kappa_{\rm D}^2\phi_q\psi_q\sinh(\kappa_{\rm D}z)w
\end{aligned}
\ee

\be
\label{A5}
\begin{aligned}
&f''-\kappa_{\rm D}^2f \\
&=-\frac{1}{2}\lambda\phi_q k''-\alpha\frac{\kappa_{\rm D}^2\phi_q}{2}k
+\alpha^2\frac{\kappa_{\rm D}^2\phi_q\psi_q}{2}\sinh(\kappa_{\rm D}z)g \\
&-(\alpha+\lambda)\psi_q\kappa_{\rm D}^2\sinh(\kappa_{\rm D} z)w
-\lambda\psi_q\kappa_{\rm D}\cosh(\kappa_{\rm D}z)w'
\end{aligned}
\ee

In addition, we need to specify the boundary conditions at $ z=\pm 0.5
$ for the four amplitude functions:
\be
\label{A6}
\begin{aligned}
w''(\pm {\textstyle \frac{1}{2}})+q_0^2w(\pm {\textstyle \frac{1}{2}})=0 \\
w'''(\pm {\textstyle \frac{1}{2}})+q_0^2 w'(\pm {\textstyle \frac{1}{2}})\mp\frac{\sigma_{t,b}}{ h} =0 \\
g''(\pm {\textstyle \frac{1}{2}})=0 ~~~ ~~~
g'''(\pm {\textstyle \frac{1}{2}})=0 \\
f(\pm {\textstyle \frac{1}{2}})=0 ~~~~~~
k(\pm {\textstyle \frac{1}{2}})=0\\
\end{aligned}
\ee




\begin{thebibliography}{99}

\bibitem{Fredrickson2006} Fredrickson, G. H. The Equilibrium Theory of Inhomogeneous Polymers.
\emph{Oxford University Press: New York} $\bf{2006}$

\bibitem{Matsen1994} Matsen, M. W.; Schick, M. Stable and unstable phases of a diblock copolymer melt.
\emph{Phys. Rev. Lett.} $\bf{1994}$, $\it{72}$, 2660-2663.

\bibitem{Bates1994} Bates, F. S.; Schulz, M. F.; Khandpur, A. K.; F\" orster, S.; Rosedale, J. H.; Almdal, K.; Mortensen, K. Fluctuations, conformational asymmetry and block copolymer phase behaviour.
\emph{Faraday Discuss.} $\bf{1994}$, $\it{98}$, 7-18.

\bibitem{Semenov1985} Semenov, A. N. Contribution to the theory of microphase layering in block-copolymer melts.
\emph{Sov. Phys. JETP} $\bf{1985}$, $\it{61}$, 733-742.

\bibitem{Leibler1980} Leibler, L. Theory of microphase separation in block copolymers.
\emph{Macromolecules} $\bf{1980}$, $\it{13}$, 1602-1617.

\bibitem{Fredrickson1987} Fredrickson, G. H.; Helfand, E. Fluctuation effects in the the
ory of microphase separation in block copolymers.
\emph{J. Chem. Phys.} $\bf{1987}$, $\it{87}$, 697-705.

\bibitem{Zschech2007} Zschech, D.; Kim, D. H.; Milenin, A. P.; Scholz, R.; Hillebrand, R.; Hawker, C. J.; Russell, T. P.; Steinhart, M.; G\" osele, U. Ordered arrays of $<100>$-oriented silicon nanorods by cmos-compatible block copolymer lithography.
\emph{Nano Lett.} $\bf{2007}$, $\it{7}$, 1516-1520.

\bibitem{Stoykovich2005} Stoykovich, M. P.; M\"uller, M.; Kim, S. O.; Solak, H. H.; Edwards, E. W.; De Pablo, J. J.; Nealey, P. F. Directed assembly of block copolymer blends into nonregular device-oriented structures.
\emph{Science} $\bf{2005}$, $\it{308}$, 1442-1446.

\bibitem{Park1997} Park, M.; Harrison, C.; Chaikin, P. M.; Register, R. A.; Adamson, D. H. Block copolymer lithography: periodic arrays of ~$10^{11}$ holes in 1 square centimeter.
\emph{Science} $\bf{1997}$, $\it{276}$, 1401-1404.

\bibitem{Thurn2000} Thurn-Albrecht, T.; Schotter, J.; K\"astle, G. A.; Emley, N.; Shibauchi, T.; Krusin-Elbaum, L.; Guarini, K.; Black, C. T.; Tuominen, M. T.; Russell, T. P. Ultrahigh-density nanowire arrays grown in self-assembled diblock copolymer templates.
\emph{Science} $\bf{2000}$, $\it{290}$, 2126-2129.

\bibitem{Amundson1991} Amundson, K.; Helfand, E.; Davis, D. D.; Quan, X.; Patel, S. S.; Smith, S. D. Effect of an electric field on block copolymer microstructure.
\emph{Macromolecules} $\bf{1991}$, $\it{24}$, 6546-6548.

\bibitem{Amundson1994} Amundson, K.; Helfand, E.; Quan, X.; Hudson, S. D.; Smith, S. D. Alignment of lamellar block copolymer microstructure in an electric field. 2. mechanisms of alignment.
\emph{Macromolecules} $\bf{1994}$, $\it{27}$, 6559-6570.

\bibitem{Morkved1996} Morkved, T. L.; Lu, M.; Urbas, A. M.; Ehrichs, E. E.; Jaeger, H. M.; Mansky, P.; Russell T. P. Local control of microdomain orientation in diblock copolymer thin films with electric fields.
\emph{Science} $\bf{1996}$, $\it{273}$, 931-933.

\bibitem{Olszowka2009} Olszowka, V.; Tsarkova, L.; B\" oker, A. Three-dimensional control over lamella orientation and order in thick block copolymer films.
\emph{Soft Matter} $\bf{2009}$, $\it{5}$, 812-819.

\bibitem{Amundson1993} Amundson, K.; Helfand, E.; Quan, X.; Smith, S. D. Alignment of lamellar block copolymer microstructure in an electric field. 1. Alignment kinetics.
\emph{Macromolecules} $\bf{1993}$, $\it{26}$, 2698-2703.

\bibitem{Majewski2012} Majewski, P. W.; Gopinadhan, M.; Osuji, C. O. J. Magnetic field alignment of block copolymers and polymer nanocomposites: Scalable microstructure control in functional soft materials.
\emph{Polym. Sci., Part B: Polym. Phys.} $\bf{2012}$, $\it{50}$, 2-8.

\bibitem{Sivaniah2003} Sivaniah, E.; Hayashi, Y.; Iino, M.; Hashimoto, T.; Fukunaga, K. Observation of perpendicular orientation in symmetric diblock copolymer thin films on rough substrates.
\emph{Macromolecules} $\bf{2003}$, $\it{36}$(16), 5894-5896.

\bibitem{Kang2012} Kang, H. M.; Craig, G. S. W.; Han, E.; Gopalan, P.; Nealey, P. F. Degree of perfection and pattern uniformity in the directed assembly of cylinder-forming block copolymer on chemically patterned surfaces.
\emph{Macromolecules} $\bf{2012}$, $\it{45}$, 159-164.

\bibitem{Man2015} Man, X. K.; Tang, J; Zhou, P; Yan, D. D.; Andelman, D. Lamellar diblock copolymers on rough substrates: self-consistent field theory studies.
\emph{Macromolecules} $\bf{2015}$, $\it{48}$, 7689-7697.

\bibitem{Man2016} Man, X.~K.; Zhou, P.; Tang, J.; Yan, D. D,; Andelman, D.
Defect-free perpendicular diblock copolymer films: the synergistic effect of surface topography and chemistry. \emph{Macromolecules} $\bf{2016}$,  $\it{49}$, 8241-8248.

\bibitem{Zhang2014} Zhang, L. S.; Wang, L. Q.; Lin, J. P. Harnessing anisotropic nanoposts to enhance long-range orientation order of directed self-assembly nanostructures via large cell simulations.
\emph{ACS Macro. Lett.} $\bf{2014}$, $\it{3}$, 712-716.

\bibitem{Zhang2016} Cong, Z. N.; Zhang, L. S.; Wang, L. Q.; Lin, J. P. Understanding the ordering mechanisms of self-assembled nanostructures of block copolymers during zone annealing.
\emph{J. Chem. Phys.} $\bf{2016}$, $\it{114}$, 114901.

\bibitem{Man2012} Man, X. K.; Andelman, D.; Orland H. Block copolymer films with free interfaces: ordering by nanopatterned substrates. \emph{Phys. Rev. E} $\bf{2012}$, $\it{86}$, 010801.

\bibitem{Man2012a} Th\'ebault, P.; Niedermayer, S.; Landis, S.; Chaix, N.; Guenoun, P.; Daillant, J.; Man, X.~K.;  Andelman, D.; Orland, H. Tailoring nanostructures using copolymer nanoimprint
         lithography. \emph{Adv. Mater.} $\bf{2012}$, $\it{24}$, 1952-1955.

\bibitem{Pujari2012} Pujari, S.; Keaton, M. A.; Chaikin, P. M.; Register, R. A. Alignment of perpendicular lamellae in block copolymer thin films by shearing.
\emph{Soft Matter} $\bf{2012}$, $\it{8}$, 5358-5363.

\bibitem{Thurn-Albrecht2000} Thurn-Albrecht, T.; DeRouchey, J.; Russell, T. P.; Jaeger, H. M. Overcoming interfacial interactions with electric fields.
\emph{Macromolecules} $\bf{2000}$, $\it{33}$, 3250-3253.

\bibitem{Xu2003} Xu, T.; Hawker, C. J.; Russell, T. P. Interfacial energy effects on the electric field alignment of symmetric diblock copolymers.
\emph{Macromolecules} $\bf{2003}$, $\it{36}$, 6178-6182.

\bibitem{Xu2004} Xu, T.; Goldbach, J. T.; Leiston-Belanger J.; Russell, T. P. Effect of ionic impurities on the electric field alignment of diblock copolymer thin films.
\emph{Colloid Polym Sci} $\bf{2004}$, $\it{282}$, 927-931.

\bibitem{Wang2006} Wang, J. Y.; Xu, T.; Leiston-Belanger, J. M.; Gupta, S.; Russell, T. P. Ion complexation: a route to enhanced block copolymer alignment with electric fields.
\emph{Phys. Rev. Lett.} $\bf{2006}$, $\it{96}$, 128301.

\bibitem{Pereira1999} Pereira, G. G.; Williams, D. R. M. Diblock copolymer melts in electric fields: the transition from parallel to perpendicular alignment using a capacitor analogy.
\emph{Macromolecules} $\bf{1999}$, $\it{32}$, 8115-8120.

\bibitem{Ashok2001} Ashok, B.; Muthukumar, M.; Russell, T. P. Confined thin film diblock copolymer in the presence of an electric field.
\emph{J. Chem. Phys.} $\bf{2001}$, $\it{115}$, 1559-1564.

\bibitem{Tsori2001} Tsori, Y.; Andelman, D. Diblock copolymer thin films: Parallel and perpendicular lamellar phases in the weak segregation limit.
\emph{Eur. Phys. J. E} $\bf{2001}$, $\it{5}$, 605-614.

\bibitem{Tsori2002} Tsori, Y.; Andelman, D. Thin film diblock copolymers in electric field: transition from perpendicular to parallel lamellae.
\emph{Macromolecules} $\bf{2002}$, $\it{35}$, 5161-5170.

\bibitem{Tsori2006} Tsori, Y.; Andelman, D.; Lin, C. Y.; Schick, M. Block copolymers in electric fields: a comparison of single-mode and self-consistent-field approximations.
\emph{Macromolecules} $\bf{2006}$, $\it{39}$, 289-293.

\bibitem{Lin2005} Lin, C. Y.; Schick, M.; Andelman, D. Structural changes of diblock copolymer melts due to an external electric field: a self-consistent-field theory study.
\emph{Macromolecules} $\bf{2005}$, $\it{38}$, 5766-5773.

\bibitem{Dehghan2015} Dehghan, A.; Schick, M.; Shi, A. C. Effect of mobile ions on the electric field needed to orient charged diblock copolymer thin films.
\emph{J. Chem. Phys.} $\bf{2015}$, $\it{143}$, 134902.

\bibitem{Tsori2003} Tsori, Y.; Tournilhac, F.; Andelman, D.; Leibler, L. Structural changes in block copolymers: coupling of electric field and mobile ions.
\emph{Phys. Rev. Lett.} $\bf{2003}$, $\it{90}$, 145504.

\bibitem{Putzel2010} Putzel, G. G.; Andelman, D.; Tsori, Y.; Schick, M. Ionic effects on the electric field needed to orient dielectric lamellae.
\emph{J. Chem. Phys.} $\bf{2010}$, $\it{132}$, 164903.



\end{thebibliography}
\end{document}